\documentclass[review, preprint,12pt]{elsarticle} %

\usepackage{amssymb}
\usepackage{amsmath,amsfonts,amssymb,bm,accents}
\usepackage{hyperref}
\usepackage{mathrsfs}
\usepackage{graphicx}
\usepackage{epstopdf}
\usepackage{float}
\usepackage{caption}
\usepackage{subcaption}
\usepackage{bbm}
\usepackage{mathrsfs}
\usepackage{cleveref}
\usepackage{soul}
\usepackage{accents}
\usepackage{graphicx}
\usepackage[export]{adjustbox}
\usepackage{xcolor}
\usepackage{courier} 
\usepackage{listings} 
\usepackage{tabu} 
\usepackage{booktabs}
\usepackage{longtable}
\usepackage{changepage} 
\usepackage[margin=2cm]{geometry}
\biboptions{sort&compress} 
\usepackage[section]{placeins}
\usepackage[font={small}]{caption}

\usepackage{siunitx}
\usepackage[export]{adjustbox}

\newcommand{\beq}{\begin{equation}}
\newcommand{\eeq}{\end{equation}}
\newcommand{\bea}{\begin{eqnarray}}
\newcommand{\eea}{\end{eqnarray}}

\newcommand{\dd}{\,{\rm d}}

\journal{Theoretical And Applied Fracture Mechanics}

\makeatletter
\def\@author#1{\g@addto@macro\elsauthors{\normalsize%
    \def\baselinestretch{1}%
    \upshape\authorsep#1\unskip\textsuperscript{%
      \ifx\@fnmark\@empty\else\unskip\sep\@fnmark\let\sep=,\fi
      \ifx\@corref\@empty\else\unskip\sep\@corref\let\sep=,\fi
      }%
    \def\authorsep{\unskip,\space}%
    \global\let\@fnmark\@empty
    \global\let\@corref\@empty  
    \global\let\sep\@empty}%
    \@eadauthor={#1}
}
\makeatother

\begin{document}

\begin{frontmatter}



\title{Phase-field chemo-mechanical modelling of corrosion-induced cracking in reinforced concrete subjected to non-uniform chloride-induced corrosion}


\author[IC]{Ev\v{z}en Korec}

\author[CTU]{Milan Jir\'{a}sek}

\author[IC]{Hong S. Wong}

\author[IC,Oxf]{Emilio Mart\'{\i}nez-Pa\~neda\corref{cor1}}
\ead{emilio.martinez-paneda@eng.ox.ac.uk}

\address[IC]{Department of Civil and Environmental Engineering, Imperial College London, London SW7 2AZ, UK}

\address[CTU]{Department of Mechanics, Faculty of Civil Engineering, Czech Technical University in Prague, Th\'{a}kurova 7, Prague - 6, 166 29, Czech Republic}

\address[Oxf]{Department of Engineering Science, University of Oxford, Oxford OX1 3PJ, UK}

\cortext[cor1]{Corresponding author.}

\begin{abstract}
A model for corrosion-induced cracking of reinforced concrete subjected to non-uniform chloride-induced corrosion is presented. The gradual corrosion initiation of the steel surface is investigated by simulating chloride transport considering binding. The transport of iron from the steel surface, its subsequent precipitation into rust, and the associated precipitation-induced pressure are explicitly modelled. Model results, obtained through finite element simulations, agree very well with experimental data, showing significantly improved accuracy over uniform corrosion modelling. The results obtained from case studies reveal that crack-facilitated transport of chlorides cannot be neglected, that the size of the anodic region must be considered, and that precipitate accumulation in pores can take years. \\ 
\end{abstract}


\end{frontmatter}


\section{Introduction}
\label{Introduction}
The objective of this study is the modelling of corrosion-induced cracking in reinforced concrete. In particular, we focus on the case when the corrosion of embedded steel is caused by chlorides. Chloride-induced and carbonation-induced corrosion are the most prevalent reasons for the corrosion-induced cracking of reinforced concrete structures and are thus responsible for the premature deterioration of 70-90\% of all concrete structures \cite{Gehlen2011-za, British_Cement_Association_BCA1997-jj}. For this reason, researchers have strived to unravel and model the mechanism of chloride-caused corrosion-induced cracking for decades. Because corrosion-induced cracking is a complex multi-physical phenomenon, time to corrosion initiation is commonly used to estimate the service life of structures, avoiding the explicit simulation of fracture. However, this approach is known to lead to overly conservative estimates because the times to cracking, spalling or delamination could be significantly longer. For this reason, the explicit modelling of processes involved in corrosion-induced fracture is necessary to obtain more realistic service life estimates.           

It has been established (see e.g. \citet{Wong2010a}) that chloride-induced corrosion results in non-uniform corrosion of the embedded steel, which leads to a non-uniformly distributed corrosion-induced pressure on the concrete. This is due to:
\begin{enumerate}[(I)]
\item The non-uniform distance of the steel surface from the chloride-contaminated concrete surface.
The closer the portion of the steel surface is to the concrete surface that is contaminated with chlorides, the earlier it is activated. 
\item Local inhomogeneities at the steel/concrete interface, which result in non-uniform corrosion initiation and pit nucleation. The exact mechanisms are still being debated \citep{Angst2019b} but, among others, metallurgical and surface defects in the steel and porosity inhomogeneities in the concrete are suspected to play a crucial role.     
\end{enumerate}
Contrarily to the well-known non-uniformity of chloride-induced corrosion, many chloride corrosion models consider only uniform corrosion, with some notable examples being the works of \citet{hansen1999a,hansen1999b} and \citet{Wei2021}. However, some researchers have recognised the necessity to consider the non-uniformity of corrosion in the modelling. For example, some models prescribed experimentally-inspired non-uniform rust-layer thickness distributions described by Gaussian \citep{Zhao2011a}, Von Mises \citep{Zhao2020b,Xi2018b} or semi-elliptical \citep{Hu2022, Fang2022} functions without explicitly resolving rust distribution or corrosion current density. More complex models, e.g. Refs. \cite{Ozbolt2012b,Geiker2017a,Chauhan2021,Chen2020}, considered the transport of chlorides and the non-uniform evolution of corrosion current density and thus the non-uniform thickness of the rust layer. Recently, phase-field fracture models have drawn the attention of the corrosion-induced cracking community due to their numerical robustness and ability to capture complex cracking phenomena of arbitrary complexity (see, e.g. \cite{Chen2021,Wu2021,Wu2022,Kristensen2021,Kristensen2020a,Kristensen2020b,Navidtehrani2022,Zhang2023,Rezaei2023,Xie2023,Zhang2023a,Han2023,Ma2023,AsurVijayaKumar2023,Costa2023,Yu2023,Min2023, Hu2023} and references therein). Certain studies modelled mesoscale fracture caused by prescribed rust layer thickness distribution \citep{Hu2022, Fang2022}; more complex ones, predicting non-uniform corrosion current density and rust layer thickness, have recently appeared, too \citep{Fang2023}. In addition, phase-field models have been recently applied to non-uniform carbonation-induced corrosion \cite{Freddi2022}.

Even though the most advanced currently available phase-field models (e.g. \cite{Fang2023}) allow capturing the gradual activation of the steel surface and the evolution of corrosion current density with remarkable complexity, the description of the formation of corrosion products and the pressure they induce on the concrete matrix is overly simplified. Corrosion products are implicitly assumed to be concentrated only in the incompressible layer separating the steel rebar from concrete. Since the concrete pores surrounding steel were observed to contain a considerable amount of corrosion products (see e.g. Ref. \cite{Wong2010a}), some models consider a porous zone or corrosion accommodation region (CAR) of a prescribed thickness around the steel rebar. Within the thickness of the porous zone, the rust layer is presumed to expand stress-free. However, it has been argued that this simplified approach has limited predictive abilities \citep{Angst2018a,Angst2019a} because it does not allow to explicitly simulate critical processes of corrosion-induced cracking, specifically: (i) the reactive transport of iron species in the pore space of concrete, (ii) the subsequent precipitation of iron species into rust that blocks the pore space, and (iii) the precipitation-induced pressure caused by the accumulation of compressible rust under confined conditions. Only recently, new models allowing the explicit simulation of processes (i)-(iii) have been proposed by \citet{Korec2023} and \citet{Pundir2023} which, contrarily to the simplified approach, allows for:
\begin{itemize}
\item Resolving the evolution of the distribution of compressible precipitates in the concrete pore space in time. Thus, the model would naturally capture the delaying effects of concrete porosity surrounding rebar and cracks on corrosion-induced cracking without the necessity to consider artificial corrosion accommodation region (CAR) around the rebar, whose size and capacity are very hard to estimate. Even if the corrosion accommodation region (CAR) would be accurately considered, it neglects the pressure of gradually forming rust in concrete porosity. This gradually increasing pressure is arguably the key driving mechanism of corrosion-induced cracking in its early stages \citep{Angst2019a, Angst2018a}. 
\item Taking into consideration the compressibility and elastic properties of the rust, which were found to significantly affect the corrosion-induced crack width \citep{Korec2023}.  
\end{itemize}

\citet{stefanoni_kinetic_2018,Furcas2022} and \citet{Zhang2021} studied the reactive transport of dissolved iron species and their precipitation in the concrete pore space, and these phenomena have been recently simulated in the mechanistic phase-field chemo-mechanical model for corrosion-induced cracking \citep{Korec2023} which explicitly models the aforementioned critical processes (i)-(iii) for the case of uniform corrosion. In this model, an eigenstrain approach is adopted to capture the pressure buildup resulting from the accumulation of precipitates, and this is coupled with a phase-field description of quasi-brittle fracture that incorporates the role of cracks in enhancing the transport of iron ions. However, phase-field fracture models for non-uniform chloride-induced corrosion considering the transport of chloride ions to steel rebars and its gradual depassivation together with processes (i)--(iii) are still missing. Thus, in this study, the model for chloride transport is incorporated into the chemo-mechanical framework (i)--(iii), implemented using the finite element method and validated with experimental results for chloride content and surface crack width.            

\section{Theory and computational framework}
\label{Sec:Theory}
In this section, let us describe the underlying theory and governing equations of the phase-field chemo-mechanical model for corrosion-induced cracking in reinforced concrete subjected to non-uniform chloride-induced corrosion. 
Following the chronological order of corrosion-induced cracking, the formulation of the theory starts in Section \ref{subSec:ReTransMod} with the reactive transport model for ionic species, i.e. chloride ions causing the gradual corrosion initiation on the steel surface and subsequently emerging ferrous and ferric ions forming rust. In Section \ref{sec:mechanics_eigenstrain}, the precipitation eigenstrain quantifying the pressure of compressible precipitates accumulating in concrete pores under confined conditions is discussed. Lastly, in Section \ref{SubSec:fractureWu}, the phase-field fracture model for quasi-brittle fracture of concrete coupled with a damage-dependant diffusivity tensor is described. The formulation of the theory is concluded with the summary of governing equations and boundary conditions in Section \ref{Sec:govEq}.\\

\noindent \emph{Notation}. Scalar quantities are denoted by light-faced italic letters, e.g. $\phi$, Cartesian vectors by upright bold letters, e.g. $\mathbf{u}$, and Cartesian second- and higher-order tensors by bold italic letters, e.g. $\bm{\sigma}$. The symbol $ \bm{1} $ represents the second-order identity tensor while $ \bm{I} $ corresponds to the fourth-order identity tensor. Inner products are denoted by a number of vertically stacked dots, where the number of dots corresponds to the number of indices over which summation takes place, such that $ \bm{\sigma}:\bm{\varepsilon} = \sigma_{ij} \varepsilon_{ij}$. Finally, $ \bm{\nabla} $ and $ \bm{\nabla} \cdot $ respectively denote the gradient and divergence operators.


\subsection{Reactive transport model}
\label{subSec:ReTransMod}
\subsubsection{Representative volume element (RVE) and primary unknown variables}
\label{subSubSec:ReTransModRVE}
For the purpose of reactive transport modelling, the porous multiphase structure of a concrete domain $\Omega^{c}$ is described by the representative volume element (RVE). Full saturation of concrete pore space is assumed and the porosity $ p_{0} = (V-V_{s})/V $ is considered to be constant in time and divided into liquid pore solution and gradually accumulating solid iron precipitates (rust), which are the product of the corrosion of embedded steel rebars.  
The content of liquid pore solution and of rust is expressed by the respective volume fractions, i.e. the liquid volume fraction $\theta_{l} = V_{l}/V$ and the precipitate volume fraction $\theta_{p} = V_{p}/V$. The portion of porosity filled by precipitates is described by the precipitate saturation ratio $ S_{p} = \theta_{p}/p_{0} $. The distribution of ionic species, i.e. free chloride ions, bound chloride ions, ferrous ions and ferric ions is described in terms of their concentrations $ c_{f} $, $ c_{b} $, $ c_{II} $ and $ c_{III} $, with units of mol per cubic meter of liquid pore solution. Let us note here that the assumption of full concrete saturation is presumed to be sensible in the close vicinity of the rebars, especially for marine structures in the tidal and splash zones. It should be emphasised that the analysis of concrete structures would require reflecting the variable water saturation caused by periodic wetting and drying cycles, which is the goal of ongoing research.

\subsubsection{Chlorides transport and corrosion initiation}
Chloride is the most common ionic form of the element chlorine, which is abundantly present in the Earth's crust and concentrates in ocean and groundwater \citep{Galan2015}. There are several ways in which chloride can contaminate concrete structures. Water containing chloride is carried away by wind above the water level and as it evaporates, a fine solid dust composed mainly of sodium chloride can be transported to considerable distances and accumulate on the surface of concrete structures. De-icing salts containing sodium chloride are commonly applied on roads in countries with colder climates and can thus easily affect bridge decks and other concrete components. Sodium chloride is also commonly used in industry and is thus abundant in wastewater, affecting the concrete parts of the sewage infrastructure. 

The transport of chlorides in concrete is mostly modelled as a diffusion process \citep{Saetta1993, KusterMaric2020a, Koenders2022}, which in certain cases allows for a closed-form solution \citep{Petcherdchoo2017}. More detailed models take into consideration the electric interaction of charged chloride with multi-ionic pore solution and advection effects \cite{Johannesson2007, Baroghel-Bouny2011}. The experimental measurements of chloride diffusivity in concrete are commonly termed ‘apparent’ because they depend on a number of conditions such as water saturation, temperature, and considered time-scale. For this reason, some modelling studies took into consideration temperature variability \citep{hansen1999a, Flint2014a} and variable saturation of concrete in real exposure conditions which undergoes regular wetting and drying cycles \citep{Ozbolt2016a, KusterMaric2020b}. Chlorides penetrate concrete not only through the concrete pore space but also through surface cracks which significantly speed up ingress \citep{Ismail2008, Djerbi2008}. The transport of chlorides through cracks was considered in the models of \citet{Savija2013} and \citet{Ozbolt2010a}.

While transported through concrete, chloride ions react with most of the phases of cement paste. It is generally agreed that the most important is the reaction with the AFm phase, $[\mathrm{Ca}_{2}(\mathrm{Al},\mathrm{Fe})(\mathrm{OH})_{6}].\mathrm{X}.\mathrm{nH}_{2}\mathrm{O}$, during which free chlorides are effectively removed from the pore solution by binding to the cement paste, forming a chloride-containing phase known as Friedel’s salt \citep{Galan2015}. Chloride binding may significantly retard the transport of chlorides through concrete and is thus considered in many models, for instance \cite{Saetta1993, hansen1999a, Johannesson2007, Baroghel-Bouny2011, KusterMaric2020a}. It appears that under certain conditions, bound chlorides can be released back to the pore solution and participate in corrosion initiation \citep{glass2000chloride, Galan2015, Angst2009}. 

Once a sufficient amount of chlorides accumulates at the steel surface, it sustains the localised breakdown of a passive semiconductive layer around the steel surface, and pit nucleation follows \citep{Poursaee2016a}. The critical amount of chlorides for corrosion initiation in reinforced concrete, known as the chloride threshold, is hard to estimate because its experimental measurements are scattered over a large range from 0.04 to 8.34$\%$ of the total chloride content by weight of cement \citep{Angst2009}, though the typical values range between 0.2 to 2.5$\%$ \citep{Glass1997}. It is generally assumed that this results from the chloride-threshold dependency on many factors, including water saturation of concrete pores surrounding steel and the properties of steel-concrete interface \citep{Angst2019b}. The corrosion current density resulting from the electrochemical reaction has been studied experimentally (e.g. \cite{Otieno2012a, Otieno2016a, Andrade2023, Walsh2016, Angst2011}), and both empirical models \citep{Otieno2016} and more complex electrochemical models \citep{Isgor2006, Mir2019, Ozbolt2017a, Geiker2017a, Chen2019b} have been proposed. The experimental studies suggest that corrosion current density during natural chloride-induced corrosion is relatively low, typically smaller than 10 \unit{\micro\ampere\per\centi\metre^2} \citep{Otieno2012a, Otieno2016a, Andrade2023, Walsh2016}. The time period until corrosion initiation, traditionally referred to as the initiation period, is followed by the propagation period during which the corrosion process actively proceeds. 

\subsubsection{Iron ion transport and precipitation}
During the propagation period, an electrochemical reaction leads to the oxidation of iron atoms on anodic sites and their subsequent release to the pore solution in the form of charged ions. Dissolved iron forms a number of complex intermediate products (see e.g. \cite{Furcas2022}), which are transported through concrete pore space by means of diffusion, electromagnetic migration and advection before finally precipitating into solid corrosion products (rust). 

Chlorides are thought to facilitate the dissolved iron transport in concrete pore solution, sharply increasing the total soluble iron concentration and thus allowing its transport and precipitation relatively far from the steel surface \citep{Glasser1989, Sagoe-Crentsil1989, Sagoe-Crentsil1993, Furcas2022}. The specific chemical mechanism is still debated \citep{Furcas2022} but it is known that at high pH of concrete, dissolved iron species form a mixed $\mathrm{Fe}^{2+}/\mathrm{Fe}^{3+}$ intermediary chloro-complex commonly known as chloride green rust \citep{Sagoe-Crentsil1993}, which is suspected from facilitating the enhanced iron transport. Consequently, studies on chloride-contaminated concrete (e.g.\ \cite{Wong2010a}) confirm that dissolved iron can be transported relatively deep into local porosity and cracks, where it gradually precipitates. 
 
Rust produced in concrete subjected to natural chloride-induced corrosion was found to be composed mostly of well-oxidised ferric hydroxyl oxides such as $\alpha-$,$\beta-$ and $\gamma-\mathrm{FeO(OH)}$ \citep{Zhang2019c}. The presence of $\beta$-FeO(OH) (Akaganeite) was reported to be particularly typical in chloride-rich concrete environments such as offshore structures \citep{ZHAO201619}. 

\FloatBarrier

\subsubsection{Chloride transport, corrosion initiation and subsequent iron ion transport and precipitation}

Based on the aforementioned current understanding, a model with the following elements is proposed. Chlorides are transported through the concrete domain $ \Omega^{c} \subset \mathbb{R}^{d}, ~ d = 2,3 $, where $ d $ is the geometrical dimension of the problem. Another domain, $\Omega^{s} \subset \mathbb{R}^{d}$, represents steel rebars. The outer concrete boundary $ \Gamma^{c} = \Gamma^{c,c} \cup \Gamma^{c,f} $ consists of two parts: $ \Gamma^{c,c}$, where the concentration of chlorides is prescribed as $ c_f = \bar{c} $, and     
$ \Gamma^{c,f} $, where the chloride flux (usually zero) is prescribed. The steel boundary or otherwise inner concrete boundary $\Gamma^{s} = \Gamma^{s,a} \cup \Gamma^{s,p} $ is divided into the active part $\Gamma^{s,a} $, where chloride threshold was reached and the corrosion process proceeds, and the passive part $ \Gamma^{s,p} $, where the corrosion process has not been initiated yet. With the ongoing transport of chlorides through concrete, $ \Gamma^{s,a} $ gradually enlarges at the expense of $ \Gamma^{s,p} $. The outward-pointing normal vector to $ \Gamma^{c} \cup \Gamma^{s} $ is $\mathbf{n}$.      
\begin{figure}[!htb]
    \centering
    \includegraphics[width=0.5\textwidth]{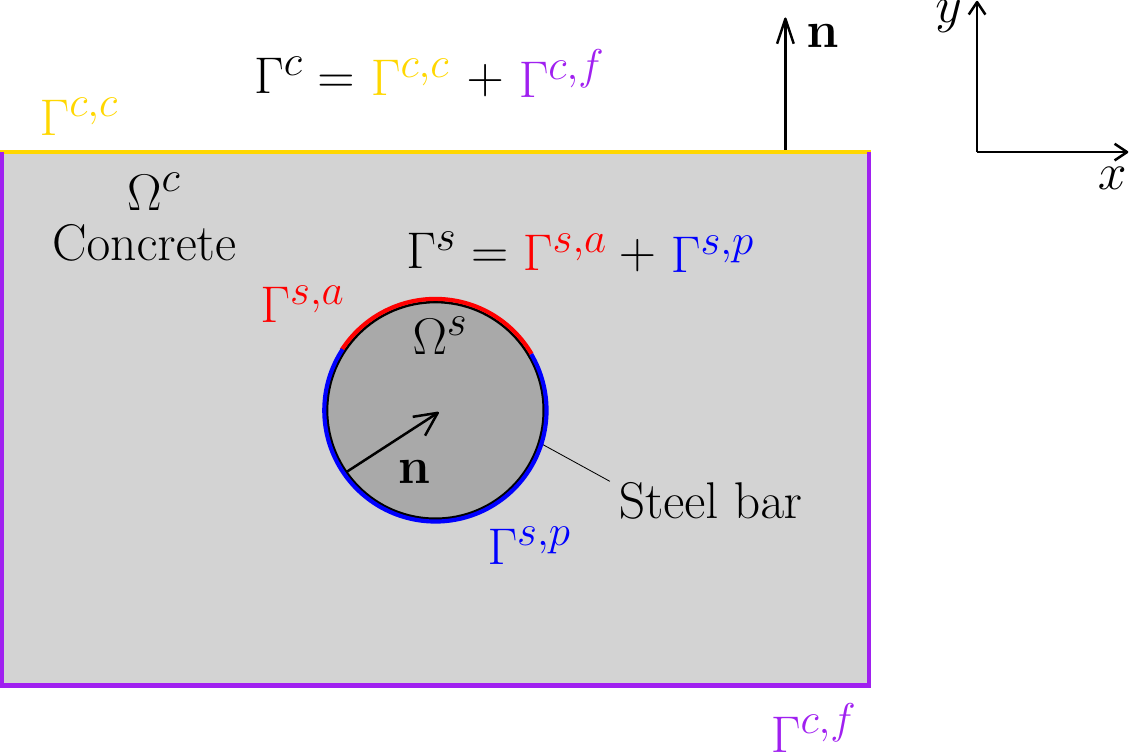}
    \caption{Graphical illustration of the domain and relevant variables for the chemo-mechanical problem.}
    \label{FigNotation}    
\end{figure}
The concentration of free chlorides $c_f$ is calculated by means of a mass-conserving transport equation derived in the study of \citet{Korec2023}. Assuming small deformations, the velocity of the solid concrete matrix is neglected. The purely diffusion-driven flux of free chlorides is considered and the flux term is scaled with liquid volume fraction following \citet{Marchand2016}. The governing equation for the transport of free chlorides reads    
\begin{equation}\label{freeCl}
\frac{\partial \left(\theta_{l}c_{f}\right)}{\partial t} - \bm{\nabla} \cdot \left(\theta_{l}\bm{D}_{f}\cdot\nabla c_{f} \right) =  - \theta_{l} R_{b} \,\,\,\,\,\,\, \text{ in } \,\, \Omega^{c},  \quad R_b = \left\langle \alpha\left(\beta c_f-c_b\right)\right\rangle 
\end{equation} 
where $\theta_{l}$ is the liquid volume fraction (see Section \ref{subSubSec:ReTransModRVE}), $\bm{D}_{f}$ is the second-order diffusivity tensor, to be defined later, and $\left\langle x\right\rangle = (x+|x|)/2$ is the positive part of $a$. Free chlorides bind to C-S-H phases, which leads to the rise of bound chloride concentration $c_b$. The rate of chloride binding $R_b = \left\langle \alpha\left(\beta c_f-c_b\right)\right\rangle$ is based on Freundlich's isotherm, assuming kinetically delayed chloride adsorption to C-S-H phases but instantaneous Friedel's salt formation as proposed by \citet{Baroghel-Bouny2011}. Freundlich's parameters $\alpha$ [1/s] and $\beta$ [1] are constants (see Table \ref{tab:tableClTranProp1}).  
Bound chlorides are considered to be immobile, their possible release from C-S-H matrix is neglected and $c_b$ is thus calculated as   
\begin{equation}\label{boundCl}
\frac{\partial \left(\theta_{l}c_{b}\right)}{\partial t} =  \theta_{l} R_{b} \,\,\,\,\,\,\, \text{ in } \,\, \Omega^{c}, \quad R_b = \left\langle\alpha\left(\beta c_f-c_b\right)\right\rangle 
\end{equation} 
Zero initial concentration is assumed for both free and bound chlorides. The corrosion process on the steel surface is initiated when the chloride threshold $T$ of locally accumulated chlorides $ C_{tot} $ is reached. In this model, $C_{tot}$ is the ratio between the total mass of chlorides (free and bound) and the mass of cement binder $m_c$ (both per unit volume). After the onset of corrosion, the corrosion current density $i_a$ locally jumps to a prescribed value $\kappa$, representing an average corrosion rate. For numerical purposes, the jump in the value of corrosion current density $i_a$ is smoothed by approximation with a polynomial function. The calculation of $ C_{tot} $ and $i_a$ is thus summarised as     
\begin{equation}\label{corrCurrThresh}
i_a= \begin{cases} 0 \,\,\,\,\,\,\, \text{ in } \,\, \Gamma^{s,p} \text{ where} & C_{tot}< T \\ \kappa \,\,\,\,\,\,\, \text{ in } \,\, \Gamma^{s,a} \text{ where} & C_{tot} \geq T \end{cases}, \quad C_{tot} = \frac{M_{Cl} p_{0}}{m_{c}}(c_{f}+c_{b})
\end{equation}
where $M_{Cl}$ is the molar mass of chlorides. 

The model for iron transport and its precipitation into rust following corrosion initiation is adopted from the study of \citet{Korec2023} that builds upon the previous works of \citet{stefanoni_kinetic_2018} and \citet{Zhang2021}. This model attempts to reduce the complexity of the underlying system of chemical reactions by considering three key processes  which facilitate the transformation of iron into rust: 
\begin{equation}\label{reaction_Fe2}
2 \mathrm{Fe}+\mathrm{O}_{2}+2 \mathrm{H}_{2} \mathrm{O} \rightarrow 2 \mathrm{Fe}^{2+}+4 \mathrm{OH}^{-}
\end{equation}
\begin{equation}\label{reaction_Fe3}
4 \mathrm{Fe}^{2+} +\mathrm{O}_{2}+2 \mathrm{H}_{2} \mathrm{O} \rightarrow 4 \mathrm{Fe}^{3+}+4 \mathrm{OH}^{-}
\end{equation}
\begin{equation}\label{reaction_FeOOH}
\mathrm{Fe}^{3+} + 3 \mathrm{OH}^{-} \rightarrow \mathrm{FeO(OH)}+\mathrm{H}_{2}\mathrm{O}
\end{equation} 
Reaction (\ref{reaction_Fe2}) describes the oxidation of iron to $ \mathrm{Fe}^{2+} $ ions on anodic sites. Ferrous ions are released from the steel surface and then transported through concrete pore solution; they are further oxidised to $ \mathrm{Fe}^{3+} $ ions by reaction (\ref{reaction_Fe3}). Eventually, reaction (\ref{reaction_FeOOH}) facilitates the precipitation of $ \mathrm{Fe}^{3+} $ ions into ferric rust.      
The precipitation of unoxidised ferrous ions is neglected, which is in line with the experimental findings of \citet{Zhang2019c} that, in naturally corroded samples subjected to chloride-induced corrosion, the content of ferrous rust is very small compared to well-oxidised ferric rust.  

The transport of ferrous and ferric ions is described by mass-conserving diffusion equations 
\begin{equation}\label{total_mass_derivative_ionic_N_2}
\frac{\partial \left(\theta_{l}c_{II}\right)}{\partial t} - \bm{\nabla} \cdot \left(\theta_{l}\bm{D}_{II}\cdot\nabla c_{II} \right) =  \theta_{l} R_{II} \,\,\ \text{ in } \,\, \Omega^{c}, \quad R_{II} = -k^{II \rightarrow III}_{r} c_{II}c_{ox}
\end{equation}  
\begin{equation}\label{total_mass_derivative_ionic_N_3}
\frac{\partial \left(\theta_{l}c_{III}\right)}{\partial t} - \bm{\nabla} \cdot \left(\theta_{l}\bm{D}_{III}\cdot\nabla c_{III} \right) =  \theta_{l} R_{III} \,\,\ \text{ in } \,\, \Omega^{c}, \quad R_{III} = -R_{II} - R_{p} = k^{II \rightarrow III}_{r} c_{II}c_{ox} - k^{III \rightarrow p}_{r} c_{III}
\end{equation}
derived in the study of \citet{Korec2023}. As in the transport of free chlorides, small deformations are assumed, the velocity of the solid concrete matrix is neglected and the flux term is scaled with liquid volume fraction, following \citet{Marchand2016}. In (\ref{total_mass_derivative_ionic_N_2}) and (\ref{total_mass_derivative_ionic_N_3}), $ \bm{D}_{II} $ and $ \bm{D}_{III} $ are the second-order diffusivity tensor of ferrous and ferric ions in pore solution, respectively. $R_{II}$, $R_{III}$ and $R_{p}$ denote the sink/source terms resulting from reactions (\ref{reaction_Fe2})--(\ref{reaction_FeOOH}).     
 
Emerging precipitates gradually fill pores and decrease the volume fraction of porosity available for transport and chemical reactions such that $ \theta_l = p_0 - \theta_p $. Precipitates are assumed to be immobile and their evolution follows the equation 
\begin{equation}\label{total_mass_derivative_crystals_N_2}
\frac{\partial \theta_{p}}{\partial t} =  \dfrac{M_{p}}{\rho_{p}} \theta_{l} R_{p} \,\,\ \text{ in } \,\, \Omega^{c}, \quad R_{p}= k^{III \rightarrow p}_{r} c_{III}
\end{equation} 
where $ M_p $ and $\rho_p$ are the molar mass and the density of precipitates, respectively. Zero initial value is assumed for $c_{II}$, $c_{III}$ and $\theta_{p}$, and equations (\ref{total_mass_derivative_ionic_N_2}) and (\ref{total_mass_derivative_ionic_N_3}) are accompanied by boundary conditions. On the actively corroding steel boundary $ \Gamma^{s,a} $, the inward influx of ferrous ions is calculated with Faraday's law as       
\begin{equation}\label{FarLaw_1}
J_{II} = \frac{i_{a}}{z_{a}F} = \mathbf{n} \cdot \boldsymbol{D}_{II} \cdot \nabla c_{II} 
\end{equation}
where $ i_{a} $ is the corrosion current density, $F$ is Faraday's constant and $ z_{a} = 2 $ represents the number of electrons exchanged in anodic reaction (\ref{reaction_Fe2}) per one atom of iron. On the remaining boundaries, zero flux of $c_{II}$ and $c_{III}$ is assumed.  

\subsection{Precipitation eigenstrain}
\label{sec:mechanics_eigenstrain}

The precipitated rust accumulates in the confined conditions of concrete pore space and has a significantly lower density than the original iron, typically by a factor of 3 to 6 \citep{Angst2019a}. Confined precipitates thus exert pressure on concrete pore walls, which is likely the key driving mechanism of corrosion-induced cracking in its early stages \citep{Angst2019a, Angst2018a}, similarly to the well-described damage mechanism of precipitating salts in porous materials \citep{Scherer1999, Flatt2014, Flatt2017a, Coussy2006, Castellazzi2013, Koniorczyk2012, Espinosa2008}. In this study, the precipitation eigenstrain model proposed by \citet{Korec2023} is employed to quantify the macroscopic stress caused by the accumulating precipitates. The small-strain tensor $\bm{\varepsilon} = \nabla_{s}\mathbf{u} = (\nabla \mathbf{u} + (\nabla \mathbf{u})^{T})/2$ is employed and it is assumed that $ \bm{\varepsilon} $ can be additively decomposed into the elastic part $ \bm{\varepsilon}_{e} $ and the precipitation eigenstrain $ \bm{\varepsilon}_{\star} $, such that $ \bm{\varepsilon} = \bm{\varepsilon}_{e} + \bm{\varepsilon}_{\star} $. The precipitation eigenstrain  is calculated as 
\begin{equation}\label{crystEigStr2}
\bm{\varepsilon}_{\star} = C S_{p}\bm{1}, \text{ with } C = \dfrac{(1-\nu)K_{p}}{(1+\nu)K_{p}+(2-4\nu)K}\left(\dfrac{\rho_{III} M_{p}}{(1-r_{0}) \rho_{p} M_{III}} - 1 \right)
\end{equation}
where $r_{0}$ is the porosity of rust and $ K_{p} = E_p/(3(1-2\nu_p)) $ is the bulk modulus of iron precipitates (rust) calculated from the Young modulus $E_p$ and the Poisson ratio $\nu_p$ of rust. Analogously, the calculation of the bulk modulus of rust-filled concrete $ K = E/(3(1-2\nu))$ requires the Young modulus $E$ and the Poisson ratio $\nu$ of rust-filled concrete. Because the mechanical properties of rust and rust-free concrete are quite different and the pores in which rust can accumulate represent a significant part of the total volume of concrete, the mechanical properties of rust-filled regions of concrete ($E,\nu$) are interpolated between the mechanical properties of rust-free concrete ($E_c$, $\nu_c$) and rust ($E_p$, $\nu_p$) based on the precipitate volume fraction $ \theta_{p} $. Adopting the rule of mixtures, the effective properties are calculated as  
\begin{equation}\label{rustFillConcProp}
E = (1-\theta_{p})E_{c} + \theta_{p}E_{p},\quad \nu = (1-\theta_{p})\nu_{c} + \theta_{p}\nu_{p}
\end{equation}

\subsection{Phase-field description of precipitation-induced cracks}
\label{SubSec:fractureWu}
Cracking of concrete is calculated with the phase-field cohesive zone model (\texttt{PF-CZM}) by Wu and co-workers \citep{Wu2017,Wu2018}. This model was chosen because it mimics the quasi-brittle behaviour of concrete by accurately capturing the softening behaviour typical of concrete-like materials. For a more detailed discussion and the derivation of the model, see Ref. \cite{Korec2023}. The steel is considered linear elastic. The primary unknown variables of the phase-field fracture problem are displacement vector $ \mathbf{u} $ and phase-field variable $ \phi $: 
\begin{equation}\label{setDispl}
\mathbf{u}(\mathbf{x},t) \in \mathbb{U} = \lbrace \forall t \geq 0: \mathbf{u}(\mathbf{x},t) \in W^{1,2}(\Omega)^{d}; \mathbf{u}(\mathbf{x},t) = \overline{\mathbf{u}}(\mathbf{x},t)\text{ on }\Gamma^{u} \rbrace
\end{equation} 
\begin{equation}\label{setPF}
\phi(\mathbf{x},t) \in \mathbb{P} = \lbrace \forall t \geq 0: \phi(\mathbf{x},t) \in W^{1,2}(\Omega^{c}); 0 \leq \phi(\mathbf{x},t) \leq 1 \text { in } \Omega^{c}; t_{1} \leq t_{2} \Longrightarrow \phi(\textbf{x},t_{1}) \leq \phi(\textbf{x},t_{2}) \rbrace
\end{equation} 
where $ \Gamma^{u} \subset (\Gamma^{c} \cup \Gamma^{s}) $ is the portion of the concrete boundary where displacements are prescribed. On the remaining part of the concrete boundary $ \Gamma^{t} = (\Gamma^{c} \cup \Gamma^{s})\setminus\Gamma^{u} $, tractions $\overline{\mathbf{t}} $ are prescribed. The volume force vector acting in domain $\Omega = \Omega^{c} + \Omega^{s}$ is denoted by $\overline{\mathbf{b}} $. The function space $ W^{1,2}(\Omega^{c})^{d} $ in (\ref{setDispl}) and (\ref{setPF}) is the Cartesian product of $ d $ Sobolev spaces $ W^{1,2}(\Omega^{c}) $ consisting of functions with square-integrable weak derivatives. The phase-field variable $ \phi $ represents the current state of damage such that $\phi=0$ denotes undamaged material and $\phi=1$ represents fully cracked material. Damage reduces the capacity of the material to carry stress and the Cauchy stress tensor $ \bm{\sigma} $ is thus given by     
\begin{equation}\label{CauchyStressTensor2}
\bm{\sigma} = g(\phi)\mathcal{\bm{C}}_{e}:(\bm{\varepsilon} - \bm{\varepsilon}_{\star}) 
\end{equation} 
where $g(\phi)$ is the degradation function characterising the remaining integrity of the material, typically a function that satisfies conditions  $ g(0) = 1 $ and $ g(1) = 0$ and is non-increasing and continuously differentiable on $[0,1]$. The appropriate choice of the degradation function serves to calibrate the model to a particular softening behaviour. $\mathcal{\bm{C}}_{e} $ is the fourth-order isotropic elastic stiffness tensor of concrete. \citet{Wu2017} proposed to express the degradation function as 
\begin{equation}\label{damFunWu1}
g(\phi)=\frac{(1-\phi)^{p}}{(1-\phi)^{p}+a_{1} \phi (1+a_{2} \phi+a_{3}  \phi^{2})}
\end{equation}
where parameters $ p \geq 2 $, $ a_{1} > 0 $, $ a_{2} $ and $ a_{3} $, allowing for the calibration of the model to particular softening behaviour, are calculated as 
\begin{equation}\label{a1a2a3}
a_{1}=\frac{4}{\pi} \frac{\ell_{irw}}{\ell}, \quad a_{2}=2 \beta_{k}^{2 / 3}-p-\frac{1}{2}, \quad a_{3}= \begin{cases} 1/2 \beta_{w}^{2}-a_{2}-1 & \text{if} \quad p=2 \\ 0 & \text{if} \quad p>2  \end{cases}
\end{equation}    
In (\ref{a1a2a3}),  $ \ell_{irw} = \widetilde{E}G_{f}/f^{2}_{t} $ is the Irwin internal length and 
\begin{equation}  
\beta_{w} = \dfrac{w_{c}}{w_{c,lin}}, \quad w_{c,lin} = \dfrac{2G_{f}}{f_{t}}
\end{equation}
\begin{equation}
\beta_{k} = \dfrac{k_{0}}{k_{0,lin}} \geq 1, \quad k_{0,lin} = -\frac{f_{t}^{2}}{2G_{f}} 
\end{equation}
where $ w_{c} $ is the limit crack opening given by the chosen softening curve and $k_{0}$ is the initial slope of the selected softening curve. $ \widetilde{E} = E(1 - \nu)/((1+\nu)(1-2 \nu))$ is the elongation modulus of concrete, $G_f$ is the fracture energy of concrete and $f_{t}$ is the tensile strength of concrete. Ratios $ \beta_{w} $ and $\beta_{k} $ compare $ w_{c} $ and $k_{0}$ with the values of the parameters governing the linear softening curve, $ w_{c,lin} $ and $k_{0,lin}$, respectively. In this model, the experimentally measured Hordijk-Cornelissen softening curve \citep{Cornelissen1962} is considered for concrete. According to \citet{Wu2017}, this type of softening is obtained by setting \begin{equation}\label{CornelissenSoft}
p = 2, \quad w_{c}=5.1361 \frac{G_{f}}{f_{t}}, \quad k_{0}=-1.3546 \frac{f_{t}^{2}}{G_{f}}
\end{equation}     
The governing equations of the deformation-fracture problem read
\begin{subequations}\label{govEqFr3}
\begin{align}
\bm{\nabla} \cdot\left(g(\phi) \mathcal{\bm{C}}_{e}: (\bm{\nabla}_{s} \mathbf{u} - \bm{\varepsilon}_{\star})\right)+\overline{\mathbf{b}} &=0 \,\,\,\,\,\,\, \text{ in } \,\, \Omega^{c} \label{govEqFr3a}\\ 
-\dfrac{1}{2}\frac{\dd g(\phi)}{\mathrm{~d} \phi} \mathcal{H}(t)+\dfrac{\ell}{\pi} G_{f} \nabla^{2} \phi-\dfrac{G_{f}}{\pi \ell}(1-\phi) &=0 \,\,\,\,\,\,\, \text{ in } \,\, \Omega^{c} \label{govEqFr3b}
\end{align}
\end{subequations} 
In (\ref{govEqFr3b}), $ \ell $ is the characteristic phase-field length scale that governs the size of the process zone \citep{Kristensen2020c}. To ensure that the results of the phase-field fracture model are length-scale insensitive, $ \ell $ has to be chosen sufficiently small such that $\ell \leq \text{min} \left( 8 \ell_{irw} / 3 \pi, \, L/100 \sim L/50 \right)$, where $L$ is the characteristic length of the structure. Also, to achieve mesh-independent results, the characteristic element length $h$ in the process zone has to be sufficiently small (5-7 times smaller than $\ell$ \citep{Kristensen2021}). The crack driving force history function $ \mathcal{H}(t) $ enforces damage irreversibility \citep{Miehe2015a} and is defined as
\begin{equation}\label{historyFunc}
\mathcal{H}(t) = \displaystyle\max_{t\in\langle 0, T \rangle} \left(\widetilde{H}, H(t)\right), \quad   \widetilde{H}=\frac{f^{2}_{t}}{2 \widetilde{E}}, \quad H(t)=\frac{\left\langle \bar{\sigma}_{1}\right\rangle^{2}}{2 \widetilde{E}}
\end{equation}
which means that $ \mathcal{H}(t) $ is the maximum value of the crack driving force $ H(t) $ that has been reached during the loading process so far, but it is at least equal to the threshold for damage nucleation, $ \widetilde{H} $. $ f_{t} $ is the tensile strength and $ \bar{\sigma}_{1} $ is the maximum principal value of the effective stress tensor $\bar{\bm{\sigma}} = \mathcal{\bm{C}}_{e}:(\bm{\varepsilon} - \bm{\varepsilon}_{\star})$ and
$ \langle\bar{\sigma}_{1}\rangle $ is its positive part. 


\subsubsection{Damage--dependent diffusivity tensor}

If cracks are filled with pore solution, they provide pathways for fast transport of dissolved ionic species compared to the surrounding concrete matrix. In this model, the fracture-induced increase of the diffusivity of dissolved species in the pore solution of concrete is captured with a damage-dependent diffusivity tensor \citep{Wu2016,Korec2023}  
\begin{equation}\label{diffusivity_tensor_2}
\theta_{l}\boldsymbol{D}_{\alpha}=\theta_{l}(1-\phi)D_{m,\alpha}\boldsymbol{1} + \phi D_{c,\alpha} \boldsymbol{1}, \quad \alpha = f, II, III
\end{equation}    
where $ D_{m,\alpha} $ is the diffusivity of the considered species in concrete and the parameter $D_{c,\alpha} \gg D_{m} $ controls the diffusivity of the cracked material.

\subsection{Overview of the governing equations}
\label{Sec:govEq}
For the sake of clarity, let us summarize the governing equations of the proposed model and the associated boundary conditions for seven unknown variables -- displacement vector $\mathbf{u}$, phase-field variable $\phi$, free chlorides concentration $c_f$, bound chlorides concentration $c_b$, 
ferrous ions concentration $c_{II}$, ferric ions concentration $c_{III}$, and precipitate volume fraction $\theta_{p}$: 
\begin{subequations}\label{summGovEq}
\begin{align}
\bm{\nabla} \cdot\left(g(\phi) \mathcal{\bm{C}}_{e}: (\bm{\nabla}_{s} \mathbf{u} - \bm{\varepsilon}_{\star})\right)+\overline{\mathbf{b}} &=0 \,\,\,\,\,\, \text{ in } \Omega^{c}\\
-\dfrac{1}{2}\frac{\dd g(\phi)}{\mathrm{~d} \phi} \mathcal{H}(t)+\dfrac{\ell}{\pi} G_{f} \nabla^{2} \phi-\dfrac{G_{f}}{\pi \ell}(1-\phi) &=0 \,\,\,\,\,\,\, \text{ in } \Omega^{c} \\
\frac{\partial \left(\theta_{l}c_{f}\right)}{\partial t} - \bm{\nabla} \cdot \left(\theta_{l}\bm{D}_{f}\cdot\nabla c_{\alpha} \right) &=  - \theta_{l} R_{b} \,\,\,\,\,\,\, \text{ in } \Omega^{c} \\
\label{summGovEq4}
\frac{\partial \left(\theta_{l}c_{b}\right)}{\partial t} &=  \theta_{l} R_{b} \,\,\,\,\,\,\, \text{ in } \Omega^{c} \\
\frac{\partial \left(\theta_{l}c_{II}\right)}{\partial t} - \bm{\nabla} \cdot \left(\theta_{l}\bm{D}_{II}\cdot\nabla c_{II} \right) &=  \theta_{l} R_{II} \,\,\,\,\,\, \text{ in } \Omega^{c} \\
\frac{\partial \left(\theta_{l}c_{III}\right)}{\partial t} - \bm{\nabla} \cdot \left(\theta_{l}\bm{D}_{III}\cdot\nabla c_{III} \right) &=  \theta_{l} R_{III} \,\,\,\,\,\, \text{ in } \Omega^{c} \\ \label{summGovEq7}
\frac{\partial \theta_{p}}{\partial t} &=  \dfrac{M_{p}}{\rho_{p}} \theta_{l} R_{p} \,\,\,\,\,\, \text{ in } \Omega^{c}
\end{align}
\end{subequations}
\begin{subequations}\label{summGovBC}
\begin{align*} 
\tag{25a}
\mathbf{u} &= \overline{\mathbf{u}} \,\,\,\,\,\,\, \text{ in } \,\,\Gamma^{u} \hspace{1cm} \bm{\sigma} \cdot \mathbf{n} = \overline{\mathbf{t}} \,\,\,\,\,\,\, \text{ in } \,\,\Gamma^{t}\\ 
\tag{25b}
\nabla \phi \cdot \mathbf{n} &= 0 \,\,\,\,\,\,\, \text{ in } \,\,\Gamma^{c} \cup \Gamma^{s}  \\
\tag{25c}
c_f &= \bar{c} \,\,\,\,\,\,\, \text{ in } \,\,\Gamma^{c,c}, \hspace{0.5cm} \hspace{0.5cm} \mathbf{n} \cdot \left(\boldsymbol{D}_{f} \cdot \nabla c_{f} \right) = 0 \,\,\,\,\,\,\, \text{ in } \,\,\Gamma^{c,f} \\
\tag{25e}
\mathbf{n} \cdot \left(\boldsymbol{D}_{II} \cdot \nabla c_{II} \right) &= 0\,\,\,\,\,\,\, \text{ in } \,\,\Gamma^{c}, \hspace{0.5cm} \hspace{0.5cm} \mathbf{n} \cdot \left(\boldsymbol{D}_{II} \cdot \nabla c_{II} \right) = \frac{i_{a}}{z_{a}F}\,\,\,\,\,\,\, \text{ in } \,\,\Gamma^{s} \\ 
\tag{25f}
\mathbf{n} \cdot \left(\boldsymbol{D}_{III} \cdot \nabla c_{III} \right) &= 0\,\,\,\,\,\,\, \text{ in } \,\,\Gamma^{c}\cup\Gamma^{s}  \\ 
\end{align*}
\end{subequations}
Boundary conditions for equations (\ref{summGovEq4}) and (\ref{summGovEq7}) are not required because these equations do not contain any space derivatives. The resulting coupled system of partial differential equations is numerically solved with the finite element method using the finite element package COMSOL Multiphysics \footnote{The COMSOL model developed is made freely available at \url{https://www.imperial.ac.uk/mechanics-materials/codes} (Note for review: it will be uploaded shortly after publication)} and the domain $\Omega^{c} \cup \Omega^{s}$ is discretised with linear triangular elements. A staggered solution scheme is employed. Direct solvers are applied to the system of linear equations in every solution step. To accelerate the solution process, significantly larger time steps (approximately 4 - 7 times) are considered in the initiation period than in the propagation period. This is because in the initiation period (i.e. before the corrosion initiation on the steel surface and thus the onset of damage), unknown variables are reduced only to free and bound chloride concentrations. 

\section{Results}
\label{Sec:Results}

The ability of the model to reproduce experimental results is validated using the experimental data reported by \citet{Chen2020} and \citet{Ye2017} for the chloride content (Section \ref{Sec:ResValCl}) and surface crack width (Section \ref{Sec:ResValCrWid}). General results observed from simulated case studies based on the reinforced mortar samples of \citet{Chen2020} and the reinforced concrete samples of \citet{Ye2017} are discussed in Section \ref{Sec:ResGenAsp} and a parametric study of chloride transport-related and corrosion activation-related parameters is provided in Section \ref{Sec:ResParStud}. The impact of varying anodic length in a three-dimensional setting is studied in Section \ref{Sec:Res3D} showcasing the potential of the model to investigate complex cases of three-dimensional non-uniform corrosion-induced cracking. 
\subsection{Choice of model parameters}
\label{Sec:modelParam}
All parameters of the proposed model have a physical basis and can be independently measured. Experimental values measured by the authors of considered case studies were thus employed where available.      
The mechanical parameters considered for the case studies are based on the tests on the reinforced mortar samples by \citet{Chen2020} and on the reinforced concrete samples by \citet{Ye2017}, and are listed in Table \ref{tab:tableMechRust1}. These two studies were chosen because they include the measurements of the surface crack width in time for naturally corroding specimens, while most experimental studies provide the surface crack width only for impressed current tests. 

The samples in both studies were prepared from Portland cement with a water-to-cement ratio of 0.6 for the samples of \citet{Chen2020}  and 0.47 for the samples of \citet{Ye2017}. While \citet{Chen2020} added only sand to the mix, \citet{Ye2017} used also crushed gravel. All concrete samples were cured for 28 days before being exposed to chlorides. \citet{Chen2020} introduced chlorides by placing the samples into a marine atmosphere environmental chamber. There, 50 g/L sodium chloride solution was regularly sprayed on the mortar surface such that the precipitation intensity of the salt solution remained approximately 5.69 g cm$^{-2}$h$^{-1}$. Chlorides were extracted by dissolving the powdered mortar samples in an acidic extraction solution. Potentiometric titration was employed to measure the chloride content. The surface crack width in time was also monitored. The specimens of \citet{Ye2017} were for 32 days subjected to cyclic 1-day wetting in a 60 g/l sodium chloride solution and 3-day oven drying. \citet{Ye2017} did not measure the chloride transport but thoroughly documented the evolution of the surface crack width in time and the steel mass loss.     

The compressive strength was provided by the authors but the remaining parameters had to be estimated. For the mortar samples of \citet{Chen2020}, the tensile strength was estimated from the porosity with the experimentally calibrated formula of \citet{Chen2013}. Young's modulus was estimated from \citet{code2022318} and the Poisson's ratio from \citet{standard2004eurocode}. For the concrete samples of \citet{Ye2017}, the tensile strength, Young's modulus and Poisson's ratio are estimated from \citet{standard2004eurocode}. The fracture energy is estimated with the formula of \citet{Bazant2002}, assuming rounded aggregates of maximum possible size for the mortar samples of \citet{Chen2020} and crushed aggregates for the samples of \citet{Ye2017}. For the steel rebar, Young's modulus of 205 GPa and Poisson's ratio of 0.28 are assumed, as these are common values for steel.\\

\begin{table}[htb!]
\begin{small}
\begin{longtable}[ht]{p{5cm} p{2.5cm} p{2cm} p{5cm}}\toprule
\multicolumn{4}{ c }{\textbf{Mechanical properties for the tests of \citet{Chen2020} and \citet{Ye2017}}} \\
\toprule
\textbf{Parameter} & \textbf{Value} & \textbf{Unit} & \textbf{Source}\\
\toprule
\toprule
Compressive strength $f_{c,cube}$ & 45.8 \& 42.5 & MPa & \cite{Chen2020,Ye2017}\\
\midrule
Tensile strength $f_{t}$ & 4.1 \& 3.2 & MPa &  \cite{Chen2013,standard2004eurocode} \\
\midrule
Young's modulus $E_{c}$ & 29 \& 34 & GPa & \cite{code2022318,standard2004eurocode}  \\
\midrule
Poisson's ratio $\nu_{c}$ & 0.18 & - & \cite{standard2004eurocode}  \\
\midrule
Fracture energy $G_{f}$ & 67 \& 100 & J m$^{-2}$ & \cite{Bazant2002} \\
\bottomrule
\caption{Model parameters: mechanical properties of concrete.}  
\label{tab:tableMechRust1}
\end{longtable}
\end{small}
\end{table}

\begin{table}[htb!]
\begin{small}
\begin{longtable}[ht]{p{6cm} p{2.5cm} p{2cm} p{5cm}}\toprule
\multicolumn{4}{ c }{\textbf{Chloride transport properties for the tests of \citet{Chen2020} and \citet{Ye2017}}} \\
\toprule
\textbf{Parameter} & \textbf{Value} & \textbf{Unit} & \textbf{Source}\\
\toprule
\toprule
Porosity $p_0$ & 0.15 \& 0.19 & - & \cite{Chen2020, Powers1946} \\
\midrule
Chloride diffusivity in undamaged concrete $\theta_{l}D_{m,f}$ & $ 2.7 \cdot 10^{-12} $  & m$^{2}$ s$^{-1}$ & \cite{stefanoni_kinetic_2018} \\
\midrule
Chloride diffusivity in cracked concrete $ D_{c,f} $ & $ 10^{-9}  $ & m$^{2}$ s$^{-1}$ &  
\cite{stefanoni_kinetic_2018} \\
\midrule
Binding isotherm parameter $\alpha$ & $10^{-5}$ & s$^{-1}$ &  \cite{Mir2019} \\
\midrule
Binding isotherm parameter $\beta$ & 0.7 & - & \cite{Mir2019}  \\
\midrule
Chloride threshold $T$ & 0.22 \& 0.56 & $\%$ of binder & \cite{Angst2009} \\
\midrule
Molar mass of chlorides $M_{Cl}$ & 35.5 & g mol$^{-1}$ & \\
\midrule
Mass fraction of cement binder $m_{c}$ & 575 \& 372 & kg m$^{-3}$ & \cite{Chen2020,Ye2017} \\
\bottomrule
\caption{Model parameters: properties related to chloride transport and corrosion initiation.}  
\label{tab:tableClTranProp1}
\end{longtable}
\end{small}
\end{table}

The parameters related to chloride transport and corrosion initiation are summarised in table \ref{tab:tableClTranProp1}. \citet{Chen2020} provided the section of porosity of mortar in the vicinity of the rebar, from which they estimated approximately 0.15 porosity of bulk mortar. The porosity of concrete for the tests of \citet{Ye2017} is estimated from the seminal work by \citet{Powers1946}, assuming that the degree of hydration is 0.9 and that only the porosity of the cement paste is relevant. The values of parameters $\alpha$ and $\beta$ of Freundlich's isotherm for chloride binding are adopted from the study of \citet{Mir2019}. \citet{Chen2020} measured the chloride profile in time and chloride diffusivity in concrete was thus obtained by a fitting procedure described in Section \ref{Sec:ResValCl}. The obtained values of chloride diffusivity are in the range reported by \citet{stefanoni_kinetic_2018}. Because \citet{Ye2017} did not measure chloride content or diffusivity, the same chloride diffusivity as for the test of \citet{Chen2020} is considered. Since no experimental measurements of chloride threshold were provided in both studies, it is estimated from the range obtained by \citet{Angst2009} by testing various values and comparing the resulting surface crack width and mass loss. The diffusivity of chlorides in fully cracked concrete is chosen the same as the diffusivity of chlorides in water.    
\begin{table}[htb!]
\begin{small}
\begin{longtable}{p{6cm} p{2cm} p{2cm} p{5cm}}
\toprule
\textbf{Parameter} & \textbf{Value} & \textbf{Unit} & \textbf{Source} \\
\toprule
\multicolumn{4}{ c }{\textbf{Properties of rust ($\bm{\mathrm{FeO(OH) + H_{2}O}}$)}} \\
\toprule
\toprule
Young's modulus $E_{p}$ & $ 440  $ & MPa & \cite{ZHAO201619} \\
\midrule
Poisson's ratio $\nu_{p}$ & $ 0.4  $ & - &  \cite{ZHAO201619} \\
\midrule
Porosity $r_{0}$ & $ 0.16 $ & - & \cite{Ansari2019} \\
\midrule
Molar mass of rust $M_{p}$ & $ 106.85  $ & g mol$^{-1}$ & \\
\midrule
Density of rust $\rho_{p}$ & $ 3560  $ & kg m$^{-3}$ & \cite{ZHAO201619} \\
\toprule
\multicolumn{4}{ c }{\textbf{Transport properties of concrete (transport of iron ions)}} \\
\toprule
\toprule
Iron ions diffusivity in undamaged concrete $\theta_{l}D_{m,II}$ and $\theta_{l}D_{m,III}$ & $ 10^{-11} $ & m$^{2}$ s$^{-1}$ & \cite{stefanoni_kinetic_2018}  \\
\midrule
Iron ions diffusivity in cracked concrete $ D_{c,II} $ and $ D_{c,III} $ & $ 7\cdot10^{-10}  $ & m$^{2}$ s$^{-1}$ &  \cite{stefanoni_kinetic_2018,Leupin2021} \\
\toprule
\multicolumn{4}{ c }{\textbf{Other chemical properties}} \\
\toprule
\toprule
Rate constant $k^{II \rightarrow III}_{r}$ & $ 0.1 $ & mol$^{-1}$m$^{3}$s$^{-1}$ & \cite{stefanoni_kinetic_2018}  \\
\midrule
Rate constant $k^{III \rightarrow p}_{r}$ & $ 2 \cdot 10^{-4}  $ & s$^{-1}$ & \cite{Leupin2021} \\
\midrule
Oxygen concentration $c_{ox}$ & $ 0.28  $ & mol m$^{-3}$ &  \cite{Zhang2021} \\
\bottomrule
\caption{Model parameters: properties of rust, transport properties of concrete and other relevant chemical properties.}
\label{tab:tableMechRust2}
\end{longtable}
\end{small}
\end{table}

The values of additional model parameters related to the properties of rust and the transport of iron ions are given in Table \ref{tab:tableMechRust2} and are identical to those considered in the study of \citet{Korec2023}, where a more detailed discussion can be found. Akaganeite ($\beta$-FeO(OH)), a commonly found rust in chloride-contaminated concrete, is assumed to be the representative corrosion product. The experimental measurements of Young's modulus and Poisson's ratio of rust are quite scattered in the literature. \citet{Korec2023} found rust elastic properties to importantly affect the surface crack width. In their study, Young's modulus and Poisson's ratio of rust were calibrated to match the surface crack width measured for the impressed current tests of \citet{Pedrosa2017}. Because corrosion was initiated by high chloride concentration in these tests, the fitted values of Young's modulus and Poisson's ratio of rust are considered relevant for this study. Overall, the chosen type and mechanical properties of rust are assumed to be valid for natural corrosion in a strongly chloride-contaminated environment.   

\FloatBarrier
\subsection{Validation of the model -- chloride transport}
\label{Sec:ResValCl}

The capability of the model to predict chloride ingress in time accurately was validated by comparing predictions with experimental data of \citet{Chen2020} at 2, 4 and 6 months. The tested mortar specimens were 300 mm long prisms with a 100 by 100 mm cross-section. Two chloride transport exposure set-ups were considered. In the first one, only the top surface of the sample was exposed to chlorides and the remaining surfaces were sealed with epoxy resin, while in the second set-up, the top surfaces and two lateral sides of the specimen were exposed with the remaining surfaces being sealed.  

\begin{figure}[!htb]
    \centering
    \begin{subfigure}[!htb]{0.49\textwidth}
    \centering
    \includegraphics[width=\textwidth]{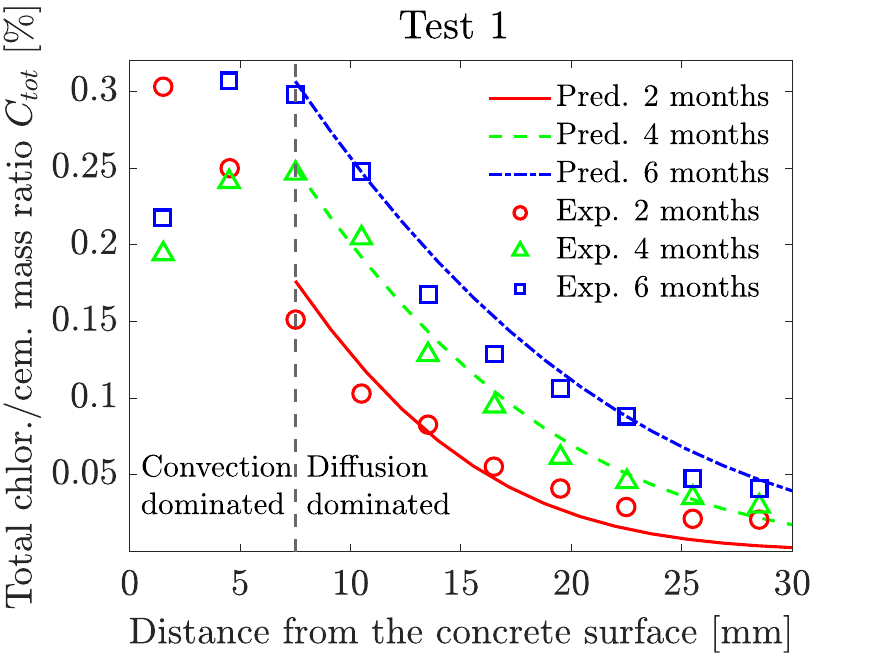}
    \caption{}
    \label{FigChenTest1}    
    \end{subfigure}
    \hfill
    \begin{subfigure}[!htb]{0.49\textwidth}
    \centering
    \includegraphics[width=\textwidth]{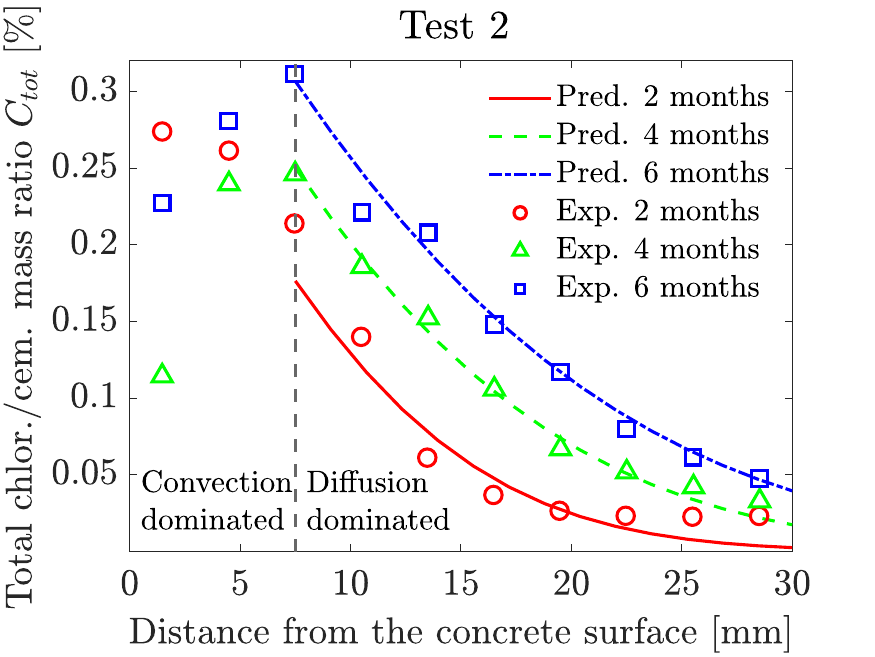}
    \caption{}
    \label{FigChenTest2} 
    \end{subfigure} 
\hfill
    \begin{subfigure}[!htb]{0.49\textwidth}
    \centering
    \includegraphics[width=\textwidth]{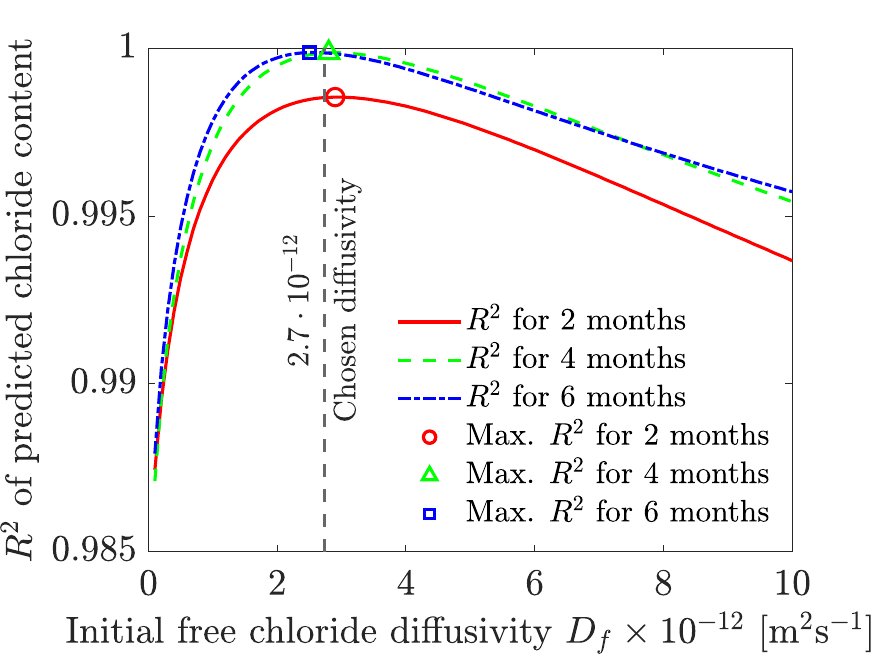}
    \caption{}
    \label{FigR2diff}    
    \end{subfigure} 
\caption{Validation of the predicted total chloride to cement mass ratio $C_{tot}$ in 2, 4 and 6 months with the experimental data from tests (a) and (b) conducted by \citet{Chen2020}. Due to the convection surface effects, the chloride content is predicted only in the diffusion-dominated region deeper than 7.5 mm. Variable concentration boundary condition based on the formula of \citet{Meira2010}, which takes into account the accumulation of precipitated salts on the concrete surface, is considered. The optimal chloride diffusion coefficient $D_f = 2.7\cdot 10^{-12} $ m$^{2}$s $^{-1}$ is obtained by comparing the coefficient of determination of the model prediction and experimental data (c) for $D_f \in \langle 10^{-13}, 10^{-11} \rangle $.}
\label{fig:ClTran}
\end{figure}

In Figs. \ref{FigChenTest1} and \ref{FigChenTest2} it can be observed that the maximum chloride concentration is not at the surface of the sample, as would be expected for a typical diffusion profile, but rather approximately 7.5 mm deeper within the sample. This behaviour results from the regular spraying of concrete with sodium chloride solution, which leads to periodic fluctuations of relative humidity in the surface layer of concrete. The distribution of chlorides in this layer is thus affected by water convection \cite{Chen2020,Meira2010} and other effects. Only deeper in the mortar where humidity remains more stable, diffusion is the dominant transport mechanism. For this reason, the transport of chlorides is modelled only in the diffusion-dominated region, which requires setting an appropriate boundary condition 7.5 mm deep in the mortar where the chloride content is maximum. \citet{Meira2010} proposed that this maximum value $ C_{max} $ can be approximated by $C_{max}=C_{0} + k_{c,max} \sqrt{D_{a c}}$ where $C_0$ is the initial chloride content, $ k_{c,max} $ is a material and environment-dependent coefficient and $ D_{ac} $ is the accumulated deposition of chlorides on the surface of concrete. $ D_{ac} $ can be calculated from the provided precipitation intensity of salt solution, and          
$ k_{c,max}  = 1.96\cdot10^{-3} $ was fitted from the provided values of chloride content 7.5 mm deep in the concrete. The optimal chloride diffusion coefficient was then found by evaluating the coefficient of determination ($R^{2}$) of the model prediction and experimental data for $D_f \in \langle 10^{-13}, 10^{-11} \rangle $. As can be seen in Fig. \ref{FigR2diff}, $R^{2}$ curves for chloride content in 2, 4 and 6 months are all concave with unique maxima around $D_f = 2.7\cdot 10^{-12} $ m$^{2}$s $^{-1}$, which was thus chosen as the optimal value. In Figs. \ref{FigChenTest1} and \ref{FigChenTest2} can be observed, the predicted chloride content agrees very well with experimental data, confirming the ability of the proposed model to simulate chloride transport accurately.       

\FloatBarrier
\subsection{Validation of the model -- mass loss and crack width}
\label{Sec:ResValCrWid}

\begin{figure}[!htb]
\begin{center}
    \begin{adjustbox}{minipage=\linewidth,scale=1}
    \centering
    \begin{subfigure}[!htb]{0.49\textwidth}
    \centering
    \includegraphics[width=\textwidth]{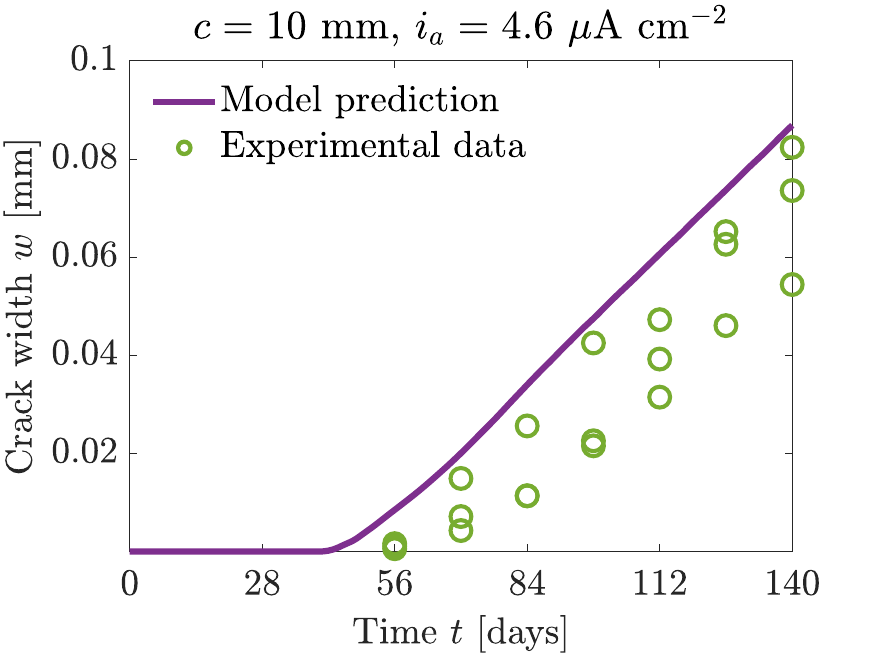}
    \caption{}
    \label{FigCrWidYe}    
    \end{subfigure}
    \hfill
    \begin{subfigure}[!htb]{0.49\textwidth}
    \centering
    \includegraphics[width=\textwidth]{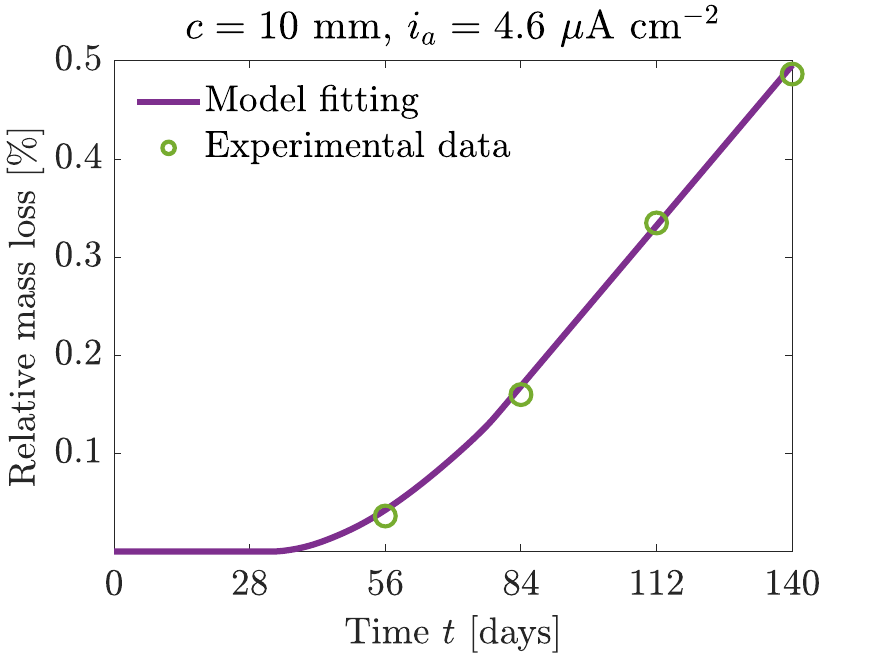}
    \caption{}
    \label{FigMassLossYe} 
    \end{subfigure} 
\end{adjustbox}
\end{center}
\caption{The predicted evolution of surface crack width (a) is compared to the experimental measurements of \citet{Ye2017} and the comparison of predicted corrosion-induced steel mass loss relative to the initial rebar mass in time and its experimental counterpart is depicted in (b). Because \citet{Ye2017} did not provide the experimental value of chloride threshold and corrosion current density, these two parameters were chosen to fit the relative mass loss curve (b). The resulting prediction of the surface crack width (a) reveals a very good agreement with experimental data, providing a safe upper bound.} 
\label{fig:CrWidYe}
\end{figure}

The capability of the model to predict the crack width in time was validated with the experimental results of \citet{Ye2017}. The tested concrete specimens with a minimal cover of 10 mm were for 32 days subjected to cyclic drying and wetting with 60 g/l sodium chloride solution. Because \citet{Ye2017} did not directly measure the parameters for chloride transport and the chloride exposure conditions were similar to the previously analysed tests of \citet{Chen2020}, the same values of parameters were employed for the tests of \citet{Ye2017}. The values of chloride threshold ($T = 0.56$ wt.$\%$ of cem.) and corrosion current density ($i_a = 4.6$ \unit{\micro\ampere\per\centi\metre^2}) were chosen by fitting the mass loss curve depicted in Fig. \ref{FigMassLossYe} to experimental data. Because the precipitation intensity of salt solution was not provided by \citet{Ye2017}, constant chloride concentration equivalent to the concentration of the sprayed 60 g/l sodium chloride solution was considered. Both fitted values lie in a range typically reported in the literature \citep{Otieno2012a, Otieno2016a, Andrade2023, Walsh2016, Angst2009}. The resulting blind prediction of the surface crack width (Fig. \ref{FigCrWidYe}) reveals a very good agreement with experimental data, providing a safe upper bound to experimental data. The predicted crack width is calculated by integration of the $x$-component of the inelastic strain tensor $ \bm{\varepsilon}_{d}  = \bm{\varepsilon} - \bm{\varepsilon}_{e} - \bm{\varepsilon}_{\star} $ ($\bm{\varepsilon}_{e}$ is the elastic part of the strain tensor) over the upper concrete surface \citep{Navidtehrani2022}:
\begin{equation}
w = \int_{\Gamma^{us}} (\bm{\varepsilon}_{d})_{x} \dd \Gamma =  \int_{\Gamma^{us}} (1-g(\phi))(\bm{\varepsilon}_{x}-(\bm{\varepsilon}_{\star})_{x}) \dd \Gamma
\end{equation}   

\FloatBarrier
\subsubsection{General aspects of the simulation results}
\label{Sec:ResGenAsp}

\begin{figure}[!htb]
    \centering
    \begin{subfigure}[!htb]{0.49\textwidth}
    \centering
    \includegraphics[width=\textwidth]{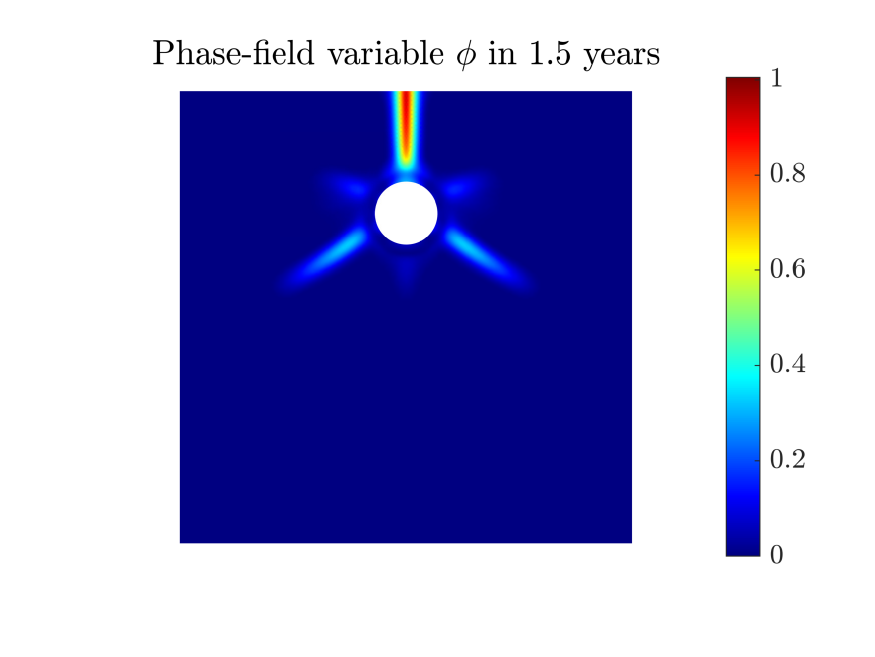}
    \caption{}
    \label{FigPF1p5years}    
    \end{subfigure}
    \hfill
    \begin{subfigure}[!htb]{0.49\textwidth}
    \centering
    \includegraphics[width=\textwidth]{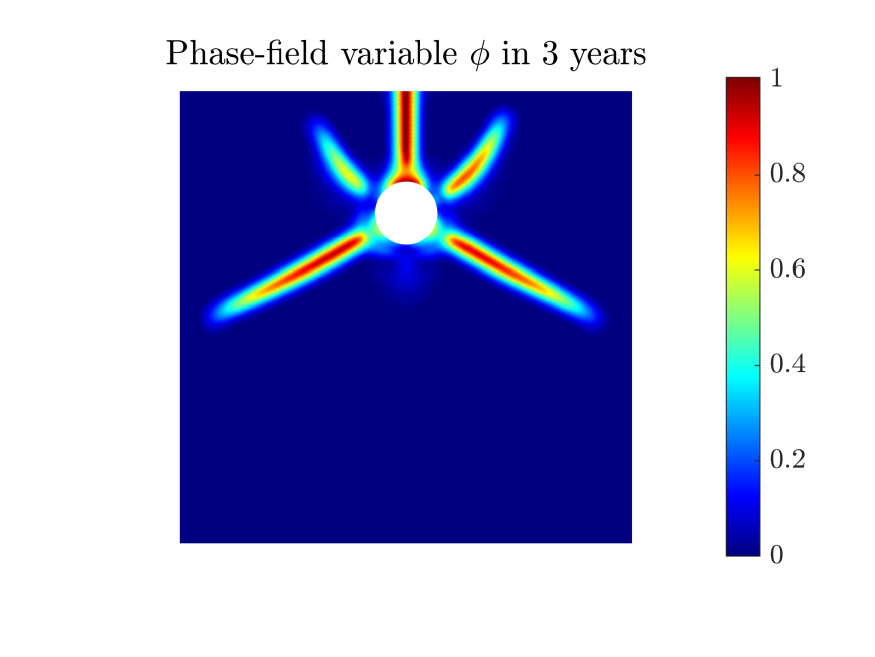}
    \caption{}
    \label{FigPF3years} 
    \end{subfigure} 
\hfill
    \begin{subfigure}[!htb]{0.49\textwidth}
    \centering
    \includegraphics[width=\textwidth]{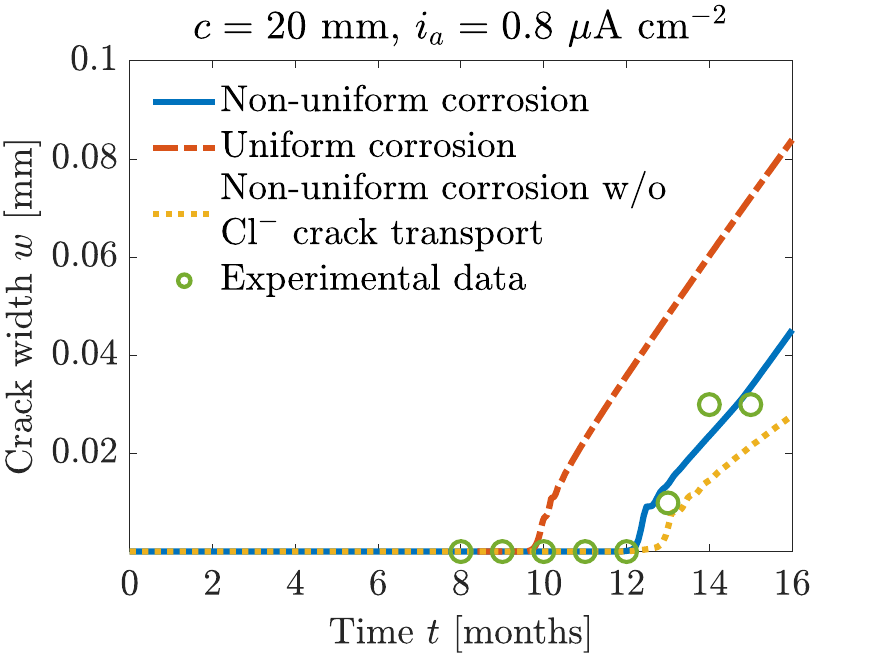}
    \caption{}
    \label{FigCrWidChen}    
    \end{subfigure} 
\caption{Predicted growth and nucleation of cracks for the work of \citet{Chen2020} characterised by the contours of the phase-field variable $\phi$ in (a) 1.5 and (b) 3 years. The evolution of surface crack width predicted by the proposed non-uniform model is compared with experimental measurements of \citet{Chen2020} (c). The comparison with alternative predictions of the uniform corrosion model and non-uniform corrosion model neglecting the transport of chloride through cracks stressing the importance of corrosion non-uniformity and crack-facilitated chloride transport is also provided in (c).}
\label{fig:PFResults}
\end{figure}

For the purposes of demonstrating the general properties of the proposed model, the simulation results obtained for the work of \citet{Chen2020}, with chlorides penetrating from the top concrete surface, are analysed in this section. The chloride threshold $T = 0.22\%$  and corrosion current density $i_a = 0.8 $ \unit{\micro\ampere\per\centi\metre^2} were chosen to fit the experimentally measured surface crack width. Cracking patterns in 1.5 and 3 years characterised by the phase-field variable are depicted in Figs. \ref{FigPF1p5years} and \ref{FigPF3years} respectively. It can be observed that the crack initially forms at the rebar surface point closest to the free concrete surface, with inclined lateral cracks forming later. The slight offset of lateral cracks from the steel surface is related to the spatial distribution of precipitates which accumulate in a thin concrete region adjacent to the steel rebar. The thin rust-filled region itself is confined between steel and remaining concrete, causing the first effective principal stress there to be initially negative, which prevents the onset of local damage. Thus, only further in the concrete, where the first principal effective stress is positive, cracks can initiate. The reader is referred to Appendix D in  \citet{Korec2023} for further context.  

\begin{figure}[!htb]
    \centering
    \begin{subfigure}[!htb]{0.49\textwidth}
    \centering
    \includegraphics[width=\textwidth]{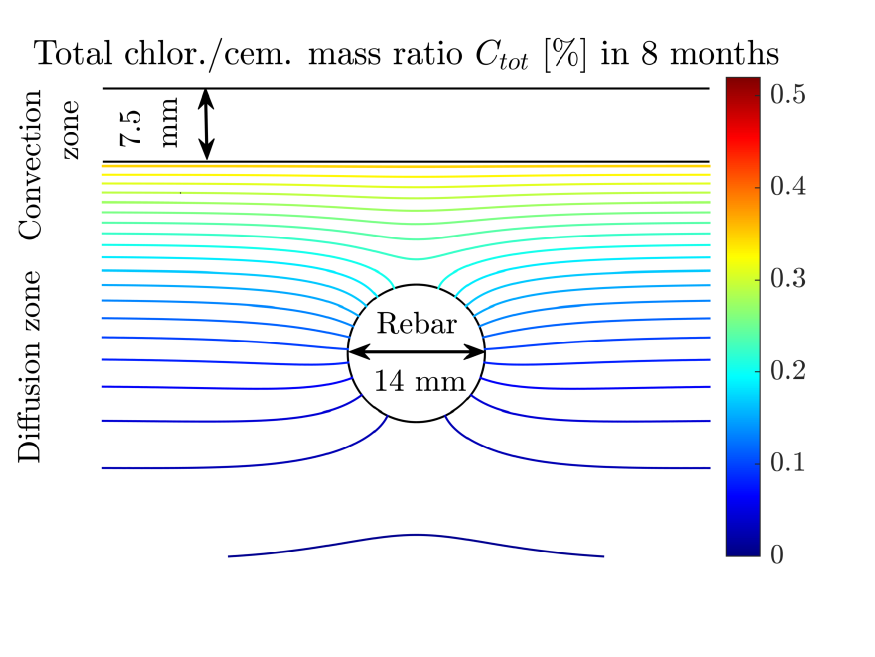}
    \caption{}
    \label{FigClTranCrc8mon}    
    \end{subfigure}
    \hfill
    \begin{subfigure}[!htb]{0.49\textwidth}
    \centering
    \includegraphics[width=\textwidth]{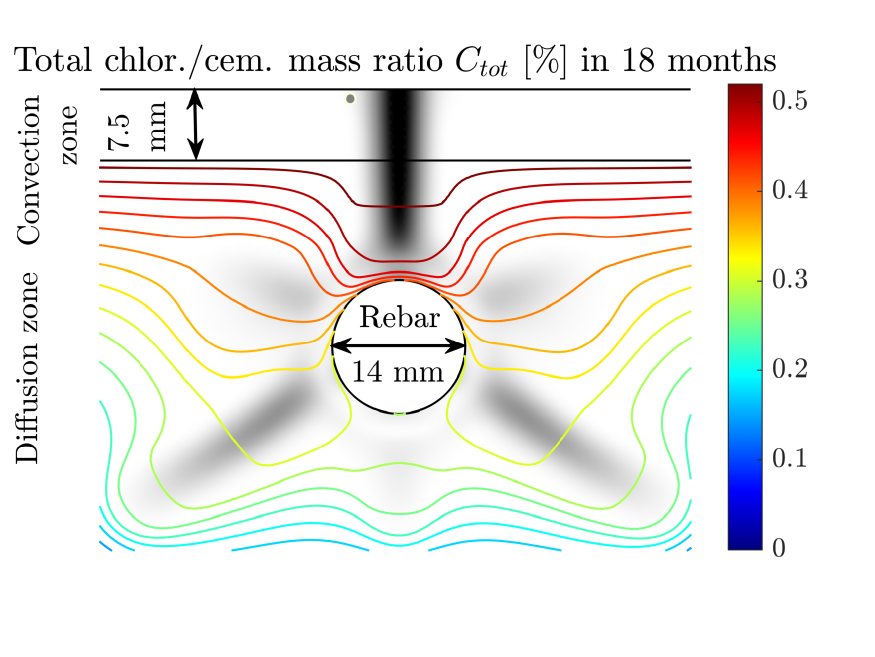}
    \caption{}
    \label{FigClTranCrc18mon} 
    \end{subfigure} 
    \hfill
\caption{The contours of total chloride content $ C_{tot} $ for the test of \citet{Chen2020} in 8 months (a) and 18 months (b) revealing the profound impact of cracks on chloride transport through concrete. The phase-field variable $\phi$ is shown in the back in a grey scale bar (0 -- white, 1 -- black).} 
\label{fig:ClTranCrc}
\end{figure}

In Fig. \ref{FigCrWidChen}, the evolution of surface crack width in time predicted by the proposed non-uniform corrosion model is compared with the prediction of two alternative models, which are simplified versions of the proposed non-uniform corrosion model. The first one is the uniform corrosion model, which considers that the entire corrosion surface starts to uniformly corrode once the penetrating chloride front reaches the chloride threshold value on the boundary of the concrete cover and steel rebar. The second alternative model considers non-uniform corrosion but neglects the enhanced transport of chlorides through corrosion-induced cracks. Comparison of crack width predicted by these three models in Fig. \ref{FigCrWidChen} reveals that although the predictions of a uniform corrosion model are on the safe side, the surface crack width is 
significantly overestimated and the vertical surface crack initiates much earlier than it should. These findings confirm the importance of considering the gradual corrosion initiation of the steel surface with the advancing chloride front. On the other hand, neglecting the enhanced transport of chlorides through cracks leads to the delay in the onset of the vertical surface crack and to underestimation of the crack width. This means that corrosion-induced cracks play an important role in chloride transport, further documented in Figs. \ref{FigClTranCrc8mon} and \ref{FigClTranCrc18mon}, and neglecting the enhanced crack-induced chloride transport is not a safe-side assumption. 

\begin{figure}[!htb]
    \begin{center}    
    \begin{subfigure}[!htb]{0.49\textwidth}
    \centering
    \includegraphics[width=\textwidth]{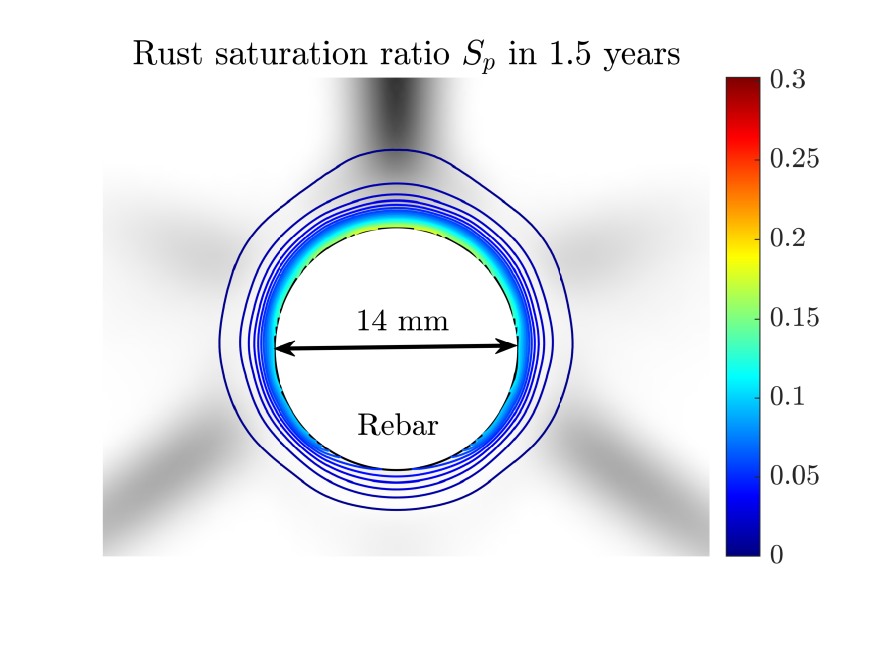}
    \caption{}
    \label{FigSp18mon}    
    \end{subfigure} 
    \hfill
    \begin{subfigure}[!htb]{0.49\textwidth}
    \centering
    \includegraphics[width=\textwidth]{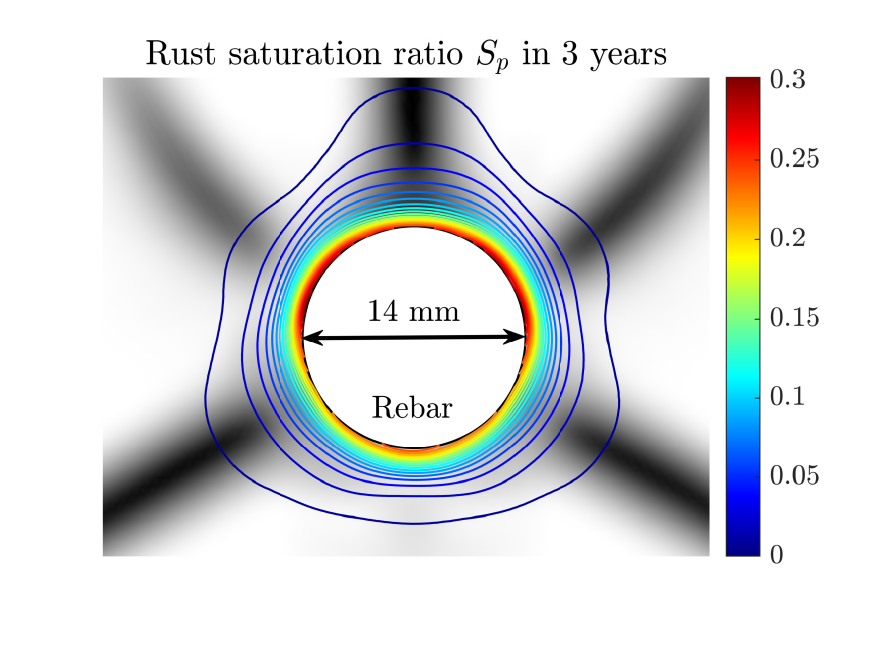}
    \caption{}
    \label{FigSp36mon}    
    \end{subfigure}
    \hfill
    \begin{subfigure}[!htb]{0.49\textwidth}
    \centering
    \includegraphics[width=\textwidth]{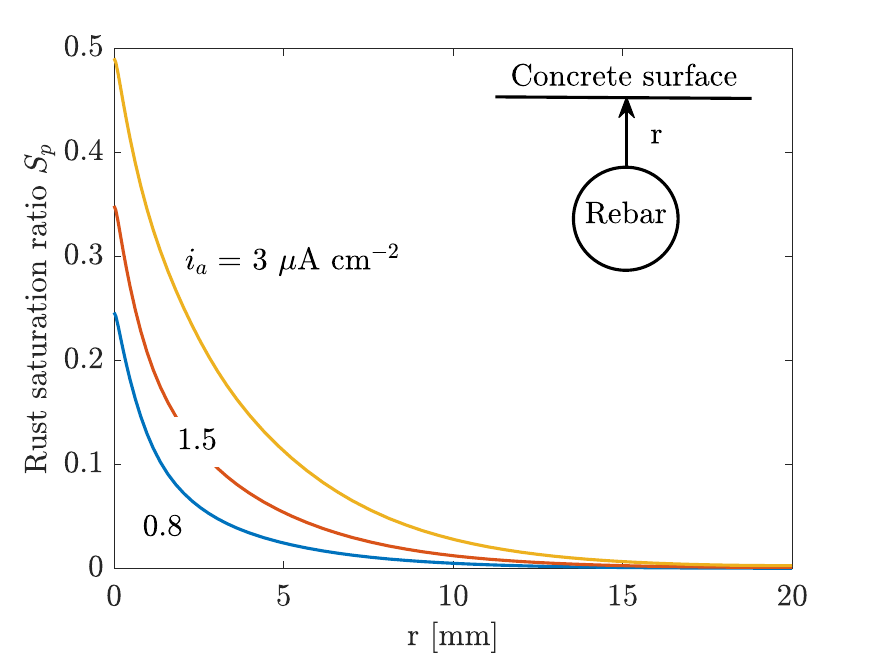}
    \caption{}
    \label{FigSpSec}    
    \end{subfigure}  
    \end{center}
    \caption{Contours of the precipitate saturation ratio $S_{p}$ in the vicinity of rebar for the test of \citet{Chen2020} in (a) 1.5 years and (b) 3 years reveal that even in three years time only $30\%$ of the concrete pore space surrounding the rebar is filled with precipitates, suggesting that the studied driving mechanism of corrosion-induced fracture, i.e., the accumulation of rust under constrained conditions in concrete pore space, could dominate for years under the low corrosion current densities of natural chloride-induced corrosion. The phase-field variable $\phi$ is shown in the back in a grey scale bar (0 -- white, 1 -- black). The comparison of the evolution of $S_{p}$ in a radial direction from the rebar ($r, \theta=0^\circ$) for corrosion current densities 0.8, 1.5 and 3 \unit{\micro\ampere\per\centi\metre^2} shows that even for values of corrosion current density that are 3.5 times higher than those considered in the simulation of the tests of \citet{Chen2020}, only less than half of the pore space is filled with precipitates in this radial section.} 
\label{fig:SpResults}
\end{figure}

A gradual corrosion initiation of the steel surface leads to a non-uniform distribution of precipitated rust, as demonstrated in Figs. \ref{FigSp18mon} and \ref{FigSp36mon} depicting the distribution of precipitates in 1.5 and 3 years, respectively. In Fig. \ref{FigSp18mon} can be observed that initially, maxima of the precipitate saturation ratio are located in the vicinity of the rebar surface closest to the concrete surface, from which chlorides penetrate. However, as depicted in Fig. \ref{FigSp36mon}, 
the distribution of precipitates in time is strongly affected by cracks that facilitate the enhanced transport of iron ions away from the steel surface (as observed experimentally by \cite{Zhu2023,Wong2010a}). For this reason, the maximum of the precipitate saturation ratio shifts to the vicinity of less-developed upper lateral cracks, which cannot facilitate the transport of as many iron ions as the other better-developed cracks. Interestingly, the predicted maximum of the precipitate saturation ratio for the tests of \citet{Chen2020} reaches only $30\%$ of the pore space. In addition, Fig. \ref{FigSpSec} shows that even if the corrosion current density were to be 3.5 times higher, only less than $50\%$ of the pore space would be filled in the considered vertical concrete section. This indicates that the considered mechanism of precipitation-induced pressure can last for years before the pore space is eventually filled. 


\FloatBarrier
\subsection{Parametric studies}
\label{Sec:ResParStud}

\begin{figure}[!htb]
    \centering
    \begin{subfigure}[!htb]{0.49\textwidth}
    \centering
    \includegraphics[width=\textwidth]{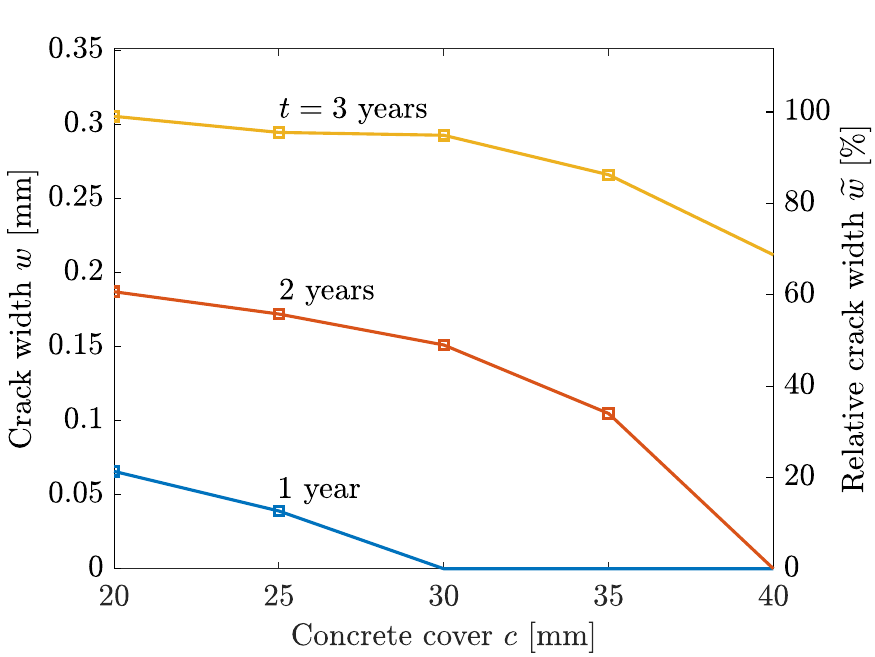}
    \caption{}
    \label{FigSweepCoverMax}    
    \end{subfigure}   
    \hfill 
    \centering
    \begin{subfigure}[!htb]{0.49\textwidth}
    \centering
    \includegraphics[width=\textwidth]{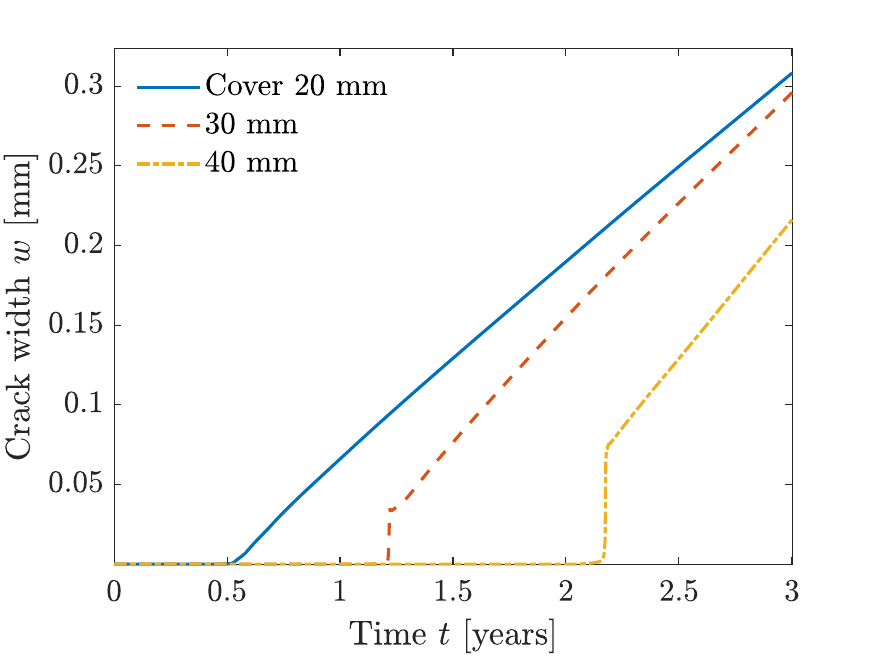}
    \caption{}
    \label{FigSweepCoverCrWid}    
    \end{subfigure}   
    \hfill 
    \centering
    \begin{subfigure}[!htb]{0.49\textwidth}
    \centering
    \includegraphics[width=\textwidth]{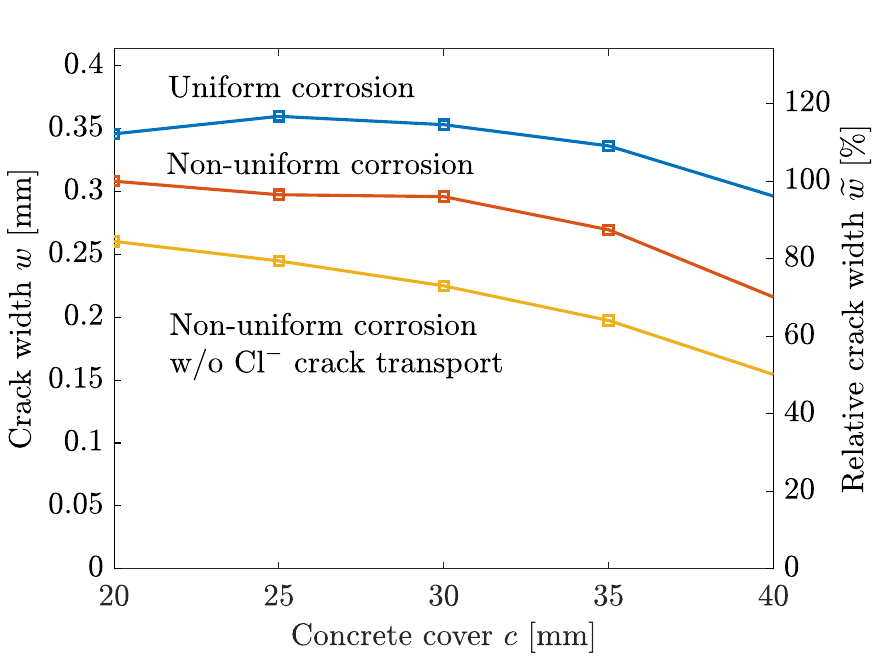}
    \caption{}
    \label{FigSweepCoverMaxComp}    
    \end{subfigure}   
\caption{Parametric study: absolute and relative surface crack width in 1, 2 and 3 years (a) and the evolution of crack width (b) for varying concrete cover. In (c), the absolute and relative crack widths in 3 years predicted by the proposed non-uniform corrosion model are compared with the alternative uniform corrosion model and non-uniform corrosion model neglecting crack-facilitated chloride transport, revealing the importance of considering non-uniform corrosion and crack-facilitated chloride transport to obtain accurate durability predictions.}
\label{fig:ResSweepCover}
\end{figure}

To better understand the impact of parameters related to chloride transport and corrosion initiation on surface crack width, parametric studies for concrete cover $ c $, chloride diffusivity in undamaged concrete $D_f$, chloride threshold $T$ and water salinity $S$ were performed. The geometry and values of the remaining model parameters were considered as in the simulated tests of \citet{Chen2020}. Chlorides penetrated only from the top concrete surface and, contrarily to the tests of \citet{Chen2020}, a constant chloride surface concentration equivalent to 35 g/l sodium chloride solution, simulating Atlantic seawater \citep{Nummelin2015, Reverdin2007}, is assumed. The simulation time was 3 years. The crack width is displayed both in its absolute value $w$ (in mm) and in its relative value $\widetilde{w}$ (in $\%$). The relative crack width $\widetilde{w}$ is calculated as the ratio of the crack width $w$ to 0.31 mm, which was the maximum crack width reached for the 20 mm cover and the parameter values are equivalent to those employed for the tests of \citet{Chen2020}. 

In Fig. \ref{FigSweepCoverMax} one can see that the increase in concrete cover leads to smaller crack widths even though the rate of crack width growth over time increases (see Fig. \ref{FigSweepCoverCrWid}). The increase of the rate of crack width with concrete cover thickness is a purely mechanical (geometric) effect, which has also been reported in computational studies of \citet{Chen2015}, \citet{Chen2020} and \citet{Korec2023}. \citet{Alonso1996} also showed experimentally that for larger covers the cracking process is delayed but the rate of the crack width in time is not smaller. However, the mechanistic increase of the rate of crack width is not able to compensate for the delaying effect that a thicker concrete cover has on the initiation of corrosion, and crack width thus decreases with increasing concrete cover. Fig. \ref{FigSweepCoverMaxComp} confirms the results demonstrated in Fig. \ref{FigCrWidChen}. It can be observed that even for varying concrete cover, the non-uniform corrosion model overestimates the predicted crack width while neglecting crack-facilitated chloride transport leads to an underestimation of the crack width.    

\begin{figure}[!htb]
\begin{center}
    \begin{adjustbox}{minipage=\linewidth,scale=0.91}
    \centering
    \begin{subfigure}[!htb]{0.49\textwidth}
    \centering
    \includegraphics[width=\textwidth]{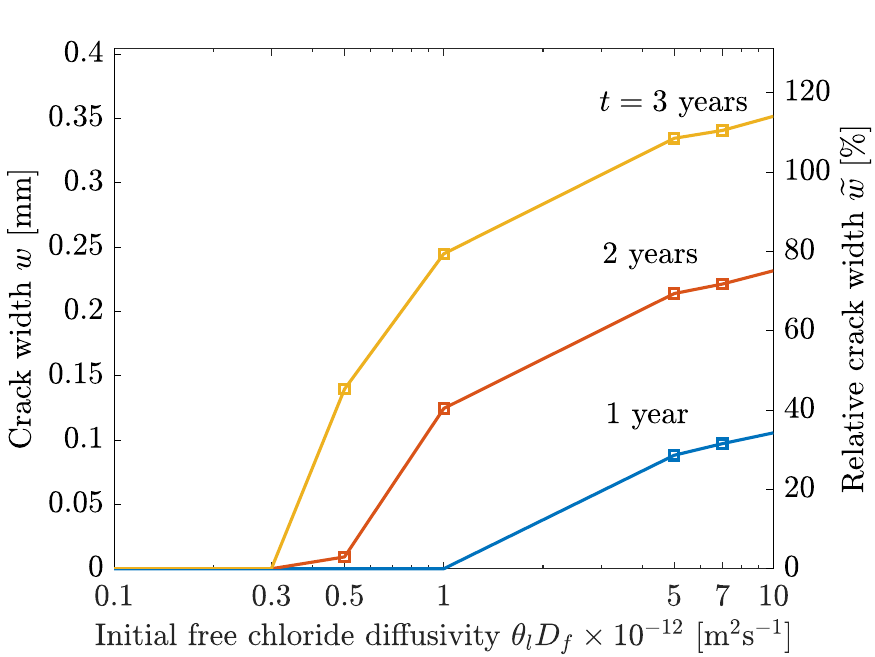}
    \caption{}
    \label{FigSweepDiffMax}    
    \end{subfigure}   
    \hfill 
    \centering
    \begin{subfigure}[!htb]{0.49\textwidth}
    \centering
    \includegraphics[width=\textwidth]{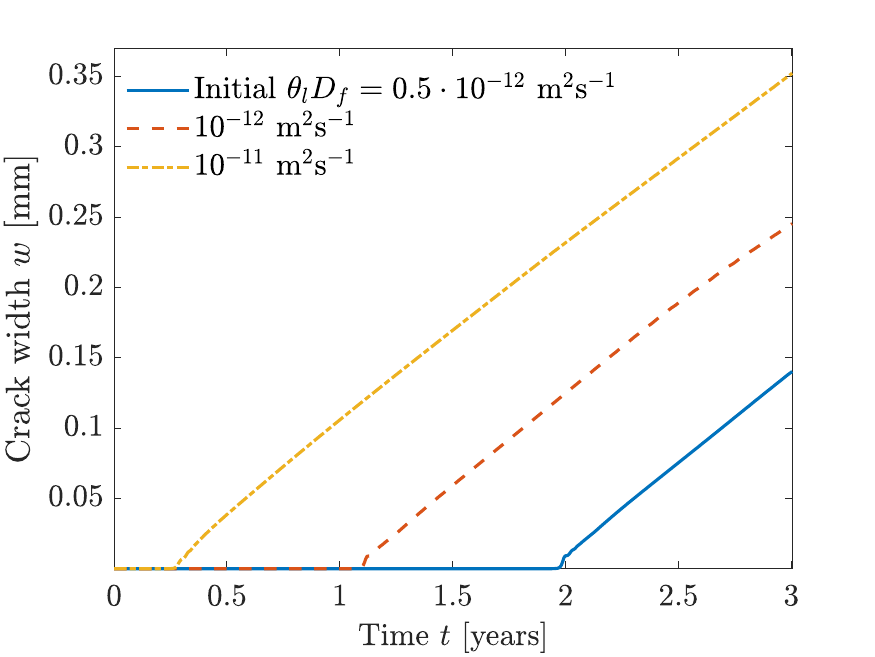}
    \caption{}
    \label{FigSweepDiffCrWid}    
    \end{subfigure}   
    \hfill 
    \centering
    \begin{subfigure}[!htb]{0.49\textwidth}
    \centering
    \includegraphics[width=\textwidth]{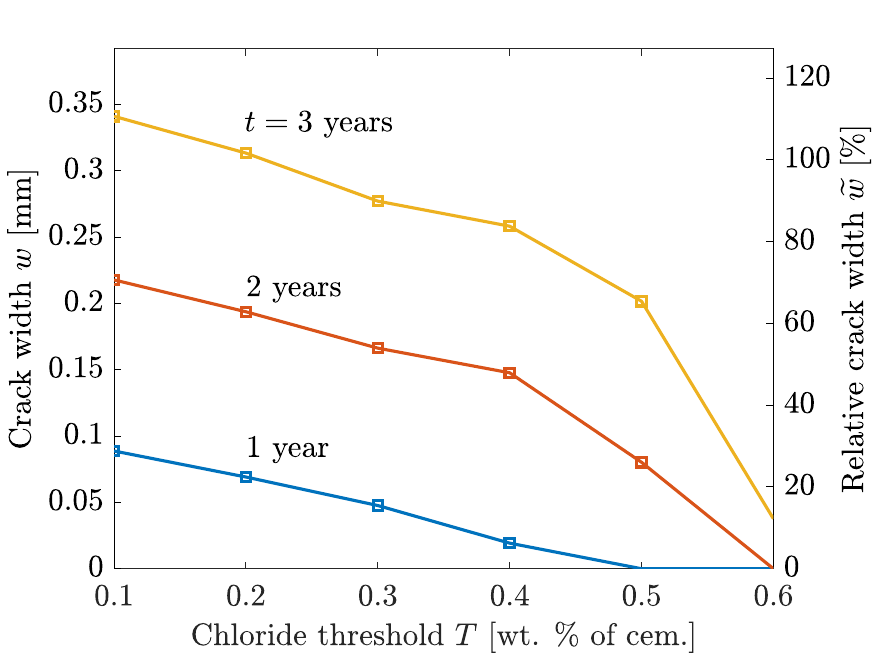}
    \caption{}
    \label{FigSweepThrMax}    
    \end{subfigure}   
    \hfill 
    \centering
    \begin{subfigure}[!htb]{0.49\textwidth}
    \centering
    \includegraphics[width=\textwidth]{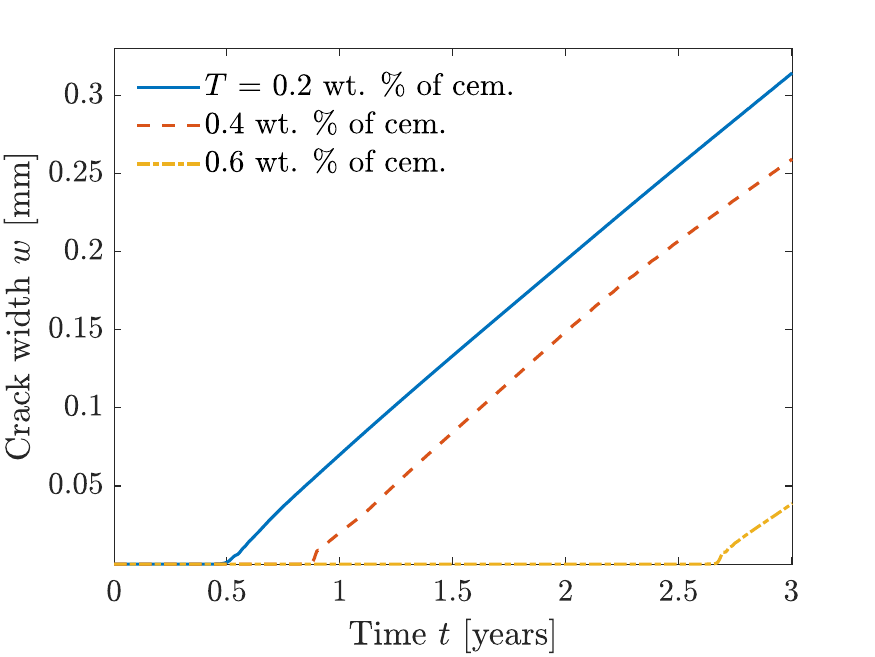}
    \caption{}
    \label{FigSweepThrCrWid}    
    \end{subfigure}   
    \hfill 
    \centering
    \begin{subfigure}[!htb]{0.49\textwidth}
    \centering
    \includegraphics[width=\textwidth]{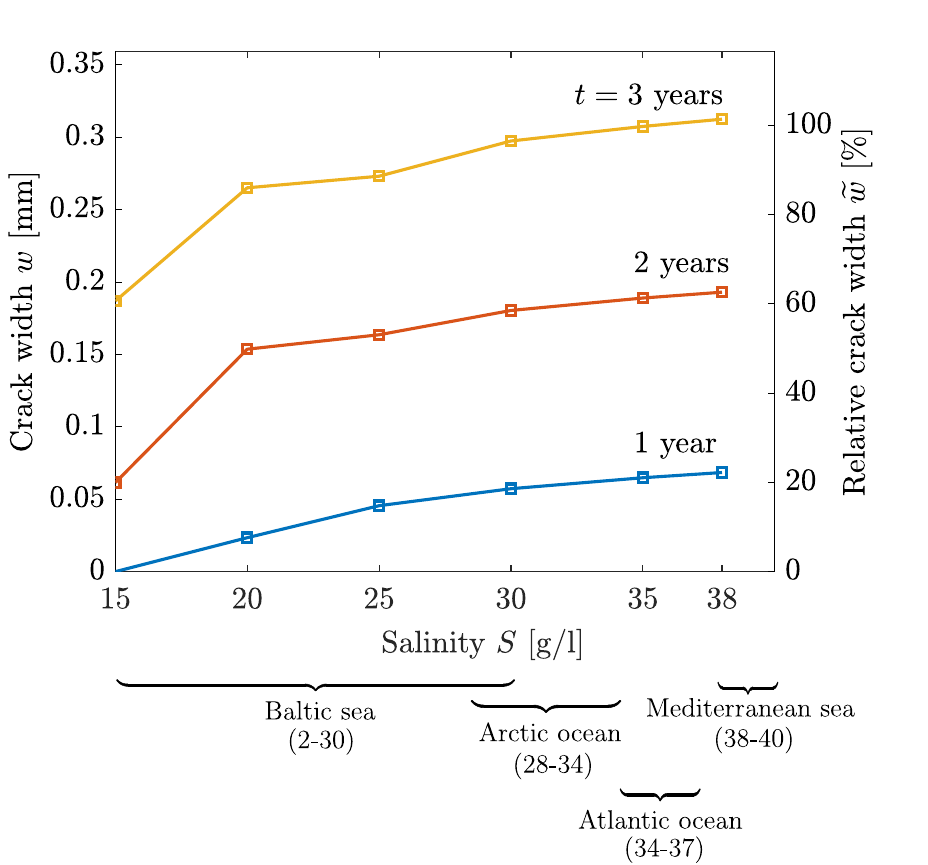}
    \caption{}
    \label{FigSweepSalMax}    
    \end{subfigure}   
    \hfill 
    \centering
    \begin{subfigure}[!htb]{0.49\textwidth}
    \centering
    \includegraphics[width=\textwidth]{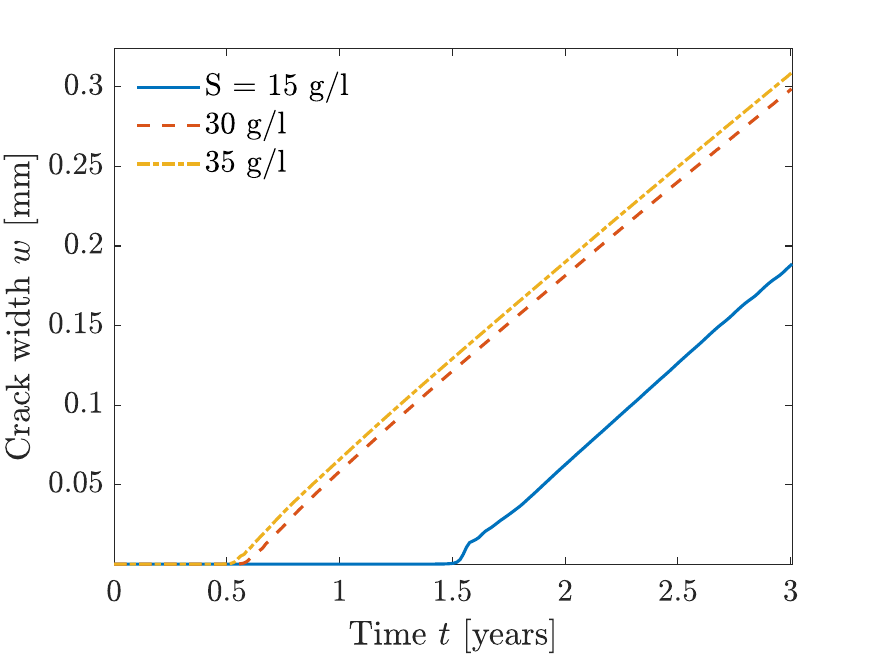}
    \caption{}
    \label{FigSweepSalCrWid}    
    \end{subfigure}   
\end{adjustbox}
\end{center}
\caption{Parametric study: absolute and relative surface crack widths in 1, 2 and 3 years on the left and the evolution of crack width on the right depending on chloride diffusivity in undamaged concrete $D_f$ (a)--(b) (with the x-axis of (a) in log scale), chloride threshold $T$ (c)--(d) and water salinity $S$ (e)--(f).}
\label{fig:ResSweep}
\end{figure}

While chloride diffusivity in undamaged concrete $D_f$ and chloride threshold $T$ have a significant impact on crack width (Figs. \ref{FigSweepDiffMax} and \ref{FigSweepThrMax}), the influence of seawater salinity $S$ (Fig. \ref{FigSweepSalMax}) is relatively low, for the values of $S$ adopted, which lie within the typical range \citep{Nummelin2015, Reverdin2007, Nessim2015, Kniebusch2019, Karna2021}. Chloride diffusivity can be reasonably measured or estimated but the profound influence of chloride threshold is troublesome because the experimental measurements are notoriously scattered in the range from 0.04 to 8.34$\%$ of total chloride content by weight of cement \citep{Angst2009}. However, this can arguably be considered more of an issue of the current state of knowledge and experimental techniques rather than a shortcoming of the model. Interestingly, variations in the value of diffusivity, chloride threshold and salinity do not lead to significant changes in the rate of crack width over time, as documented in Figs. \ref{FigSweepDiffCrWid}, \ref{FigSweepThrCrWid} and \ref{FigSweepSalCrWid}.    
 
\FloatBarrier
\subsection{Non-uniform corrosion in 3D}
\label{Sec:Res3D}

\begin{figure}[!htb]
    \centering
    \begin{subfigure}[!htb]{0.49\textwidth}
    \centering
    \includegraphics[width=\textwidth]{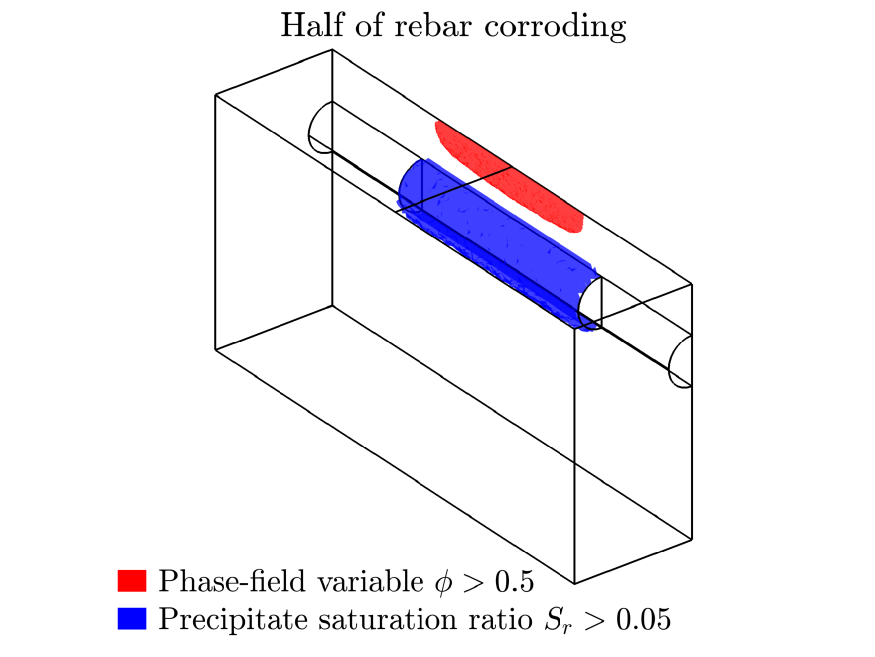}
    \caption{}
    \label{FigRes3Dhalf}    
    \end{subfigure}   
    \hfill 
    \centering
    \begin{subfigure}[!htb]{0.49\textwidth}
    \centering
    \includegraphics[width=\textwidth]{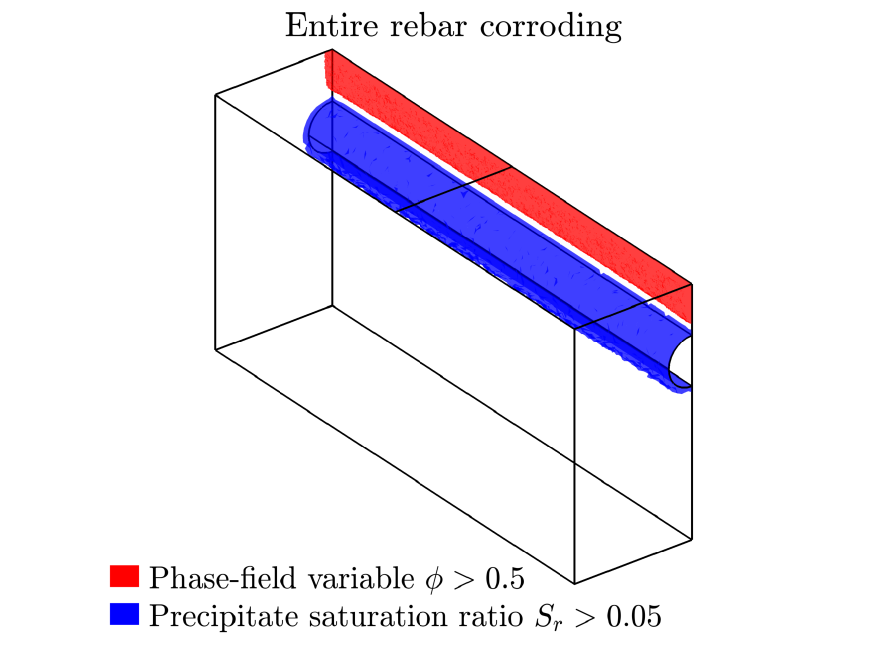}
    \caption{}
    \label{FigRes3Dwhole}    
    \end{subfigure}   
    \hfill 
    \centering
    \begin{subfigure}[!htb]{0.49\textwidth}
    \centering
    \includegraphics[width=\textwidth]{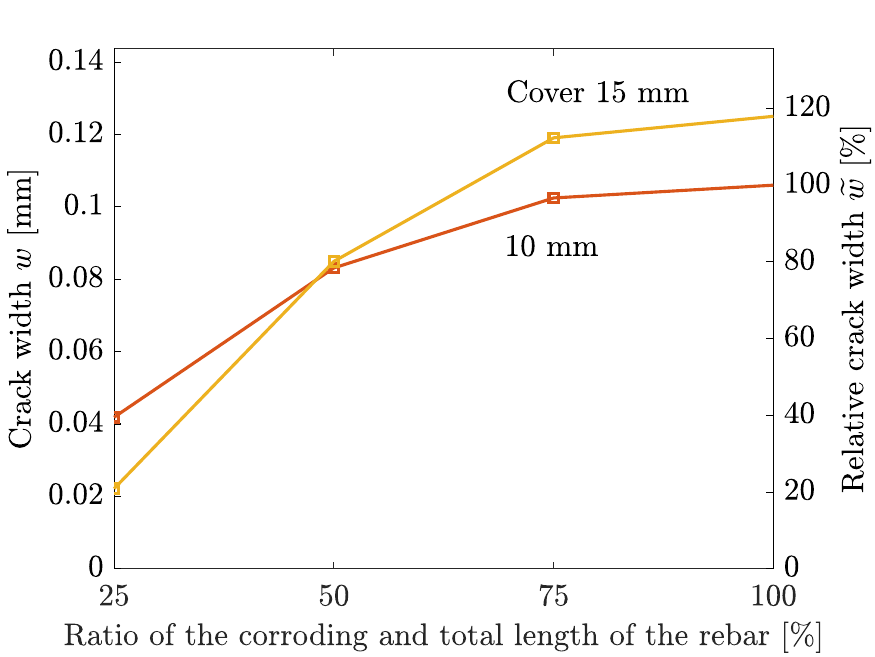}
    \caption{}
    \label{Fig3DcrWid}    
    \end{subfigure}   
\caption{Contours of evolving cracks characterised by contours of the phase-field (in red) and corroding rebar region represented by $ S_p > 0.05 $ (in blue), for (a) $50\%$ of rebar surface uniformly corroding, (b) $100\%$ of rebar surface uniformly corroding (both samples are halved). Absolute and relative surface crack widths in the middle of the sample in 60 days for a varying length of the anodic region and covers 10 and 15 mm are depicted in (c).}
\label{fig:Res3D}
\end{figure}

In the previous sections, only two-dimensional boundary value problems were analysed, using the implicit assumption that corrosion proceeds uniformly over the length of the rebar. This is realistic for well-controlled tests in laboratory conditions, on which the proposed model was validated. However, chloride-induced corrosion in real concrete structures could be significantly more localised. This may result for example from additional load-induced, thermal-induced or shrinkage-induced cracks or simply from material heterogeneities or construction imperfections. To better understand the impact of varying anodic length on the surface crack width,  a 100 mm long prism with a 50 by 50 mm cross-section was simulated as a three-dimensional domain. A rebar of 10 mm in diameter and concrete covers of 10 and 15 mm were considered. The sample was corroded for 60 days with the corrosion current density of 6 \unit{\micro\ampere\per\centi\metre^2} uniformly distributed over the anodic region (see Figs. \ref{FigRes3Dhalf} and \ref{FigRes3Dwhole}). The uniform distribution of corrosion current density was introduced to facilitate the interpretation of the results. The parametric study for different ratios of corroding (anodic) and total length of the rebar (Fig. \ref{Fig3DcrWid}) revealed a strong influence of the studied ratio on the surface crack width. This agrees with the experimental findings of \citet{Torres-Acosta2004} who identified a strong impact of the length of the anodic region on the time to the appearance of a surface crack. 

Because of the high sensitivity of the surface crack width to the anodic length depicted in Fig. \ref{Fig3DcrWid}, a heavily corroded small section of a rebar may lead to a similar crack width as a much less corroded larger rebar region. Evaluation of the state of corrosion only from the surface crack width may lead to a catastrophically erroneous assessment of ultimate strength capacity because the surface crack width cannot reveal if the rebar is not literally disconnected by corrosion and thus can no longer fulfil its load-bearing function properly. From an experimental perspective, similar conclusions were drawn by \citet{Andrade2016a}. The obtained results also indicate that two-dimensional simulations implicitly assuming uniformity of corrosion in length cannot be straightforwardly extended to real structures, and three-dimensional simulations assuming possible variations in the length of anodic regions are necessary. The proposed model is capable of analysing sections of complex reinforced concrete components and can be thus employed to analyse various academically and industrially relevant scenarios of highly localised chloride corrosion.   

The proposed model allows the simulation of both the initiation and propagation stages of chloride-induced corrosion in reinforced concrete structures. The necessary inputs are chloride diffusivity, chloride binding isotherm parameters, chloride threshold and concrete porosity, tensile strength, Young's modulus, Poisson's ratio and fracture energy. Also, the model takes the value of the corrosion current density as a parameter. One has to bear in mind that the corrosion current density is strongly affected among other things by the water saturation of porosity \citep{Stefanoni2019} which is variable during the year depending on exposure conditions \citep{Andrade1999}. 

Full saturation of concrete is assumed, which is sensible in the close vicinity of the rebars for many structures, including marine structures in the tidal and splash zones. However, because concrete structures exposed to the atmosphere commonly undergo wetting and drying cycles, coupling with an external water transport model is recommended for long-term structural analysis. In this case, the impact of creep on the mechanical properties of concrete should also be considered.

\section{Conclusions}
\label{Sec:Conclusions}

In this study, a model for corrosion-induced cracking of reinforced concrete subjected to chloride-induced corrosion was presented. It consists of seven differential equations for seven associated unknown field variables, which are solved by the finite element method. The ability of the proposed model to simulate both chloride transport and corrosion-induced cracking accurately was validated against the experimental results of \citet{Chen2020} and \citet{Ye2017}, revealing very good agreement with the experimentally measured chloride content and crack width in time. The main findings observed in the simulated case studies are:
\begin{itemize}
\item For the case of chloride-induced corrosion, the proposed non-uniform model provides significantly more accurate crack width estimates than the simplified uniform corrosion model.  
\item Neglecting crack-facilitated chloride transport is not a conservative assumption, as it leads to an underestimation of the crack width.
\item The numerical results from analysed case studies suggest that under the conditions of natural chloride-induced corrosion, where the current density is typically below 10 \unit{\micro\ampere\per\centi\metre^2}, the considered mechanism of precipitation-induced pressure can last for years before pore space is eventually filled.  
\item Variations of the values of parameters related to chloride transport and corrosion initiation, namely chloride diffusivity, seawater salinity and chloride threshold change the crack initiation time but the rate of crack width growth is not affected significantly. 
\item Varying length of the corroding (anodic) region on the rebar surface significantly affects the crack width. Because chloride-induced corrosion is well-known to have a pitting character, the obtained results indicate that it is not possible to draw a direct link between the mass loss of steel rebars and the surface crack width, unless the distribution and size of anodic regions are known. 
\end{itemize}     
There are many opportunities for future research; for example, coupling the proposed chemo-mechanical model with the electrochemical corrosion model to predict corrosion current density, or extending the model to combined carbonation-induced and chloride-induced corrosion.    

\section{Acknowledgements}
\label{Acknowledge of funding}

The authors gratefully acknowledge stimulating discussions with Prof Nick Buenfeld (Imperial College London), Prof Carmen Andrade (CIMNE), Jiahang Yu (Imperial College London) and Prof Milan Kouril (University of Chemistry and Technology, Prague). E. Korec acknowledges financial support from the Imperial College President’s PhD Scholarships. M. Jirásek acknowledges the support of the European Regional Development Fund (Center of Advanced Applied Sciences, project CZ.02.1.01/0.0/0.0/16\_19/0000778). E. Mart\'{\i}nez-Pa\~neda was supported by an UKRI Future Leaders Fellowship [grant MR/V024124/1]. We additionally acknowledge computational resources and support provided by the Imperial College Research Computing Service (http://doi.org/10.14469/hpc/2232).



\end{document}